\documentclass[aps,twocolumn,superscriptaddress,prd]{revtex4-1}

\usepackage{bm,graphicx,textcomp,amssymb,amsmath,dcolumn,color}
  
\usepackage{multirow}
\usepackage{tabularx}
\usepackage{ctable} 
\usepackage{siunitx}   
\usepackage[utf8]{inputenc}
\usepackage[T1]{fontenc}

\usepackage{bbm}
\usepackage{bbold}   
\usepackage{mathtools}
\newcommand*{\I}{ {\mathrm i} }
\newcommand*{\ee}{ {\mathrm e} }
 
\usepackage{ulem}
\usepackage{comment}

\usepackage{lipsum}  
\usepackage{graphicx,stackengine}
\newcommand\underlay[4]{%
  \stackengine{0pt}%
  {\kern#2\includegraphics[height=#1]{#4}}%
  {\includegraphics[height=#1]{#3}}%
  {O}{l}{F}{F}{L}%
}
\newcommand\addunderlay[4]{%
  \stackengine{0pt}%
  {\kern#2\includegraphics[height=#1]{#4}}%
  {#3}%
  {O}{l}{F}{F}{L}%
}

\begin{document}

\title{Dynamically Assisted Tunneling in the Floquet Picture} 

\author{Daniil Ryndyk}
\affiliation{Helmholtz-Zentrum Dresden-Rossendorf, 
Bautzner Landstra{\ss}e 400, 01328 Dresden, Germany,}

\author{Christian Kohlf\"urst}
\affiliation{Helmholtz-Zentrum Dresden-Rossendorf, 
Bautzner Landstra{\ss}e 400, 01328 Dresden, Germany,}

\author{Friedemann Queisser}
\affiliation{Helmholtz-Zentrum Dresden-Rossendorf, 
Bautzner Landstra{\ss}e 400, 01328 Dresden, Germany,}

\author{Ralf Sch\"utzhold}
\affiliation{Helmholtz-Zentrum Dresden-Rossendorf, 
Bautzner Landstra{\ss}e 400, 01328 Dresden, Germany,}

\affiliation{Institut f\"ur Theoretische Physik, 
Technische Universit\"at Dresden, 01062 Dresden, Germany,}

\date{\today}

\begin{abstract}
We study how tunneling through a potential barrier $V(x)$ can be enhanced 
by an additional harmonically oscillating electric field 
${\mathfrak E}(t)={\mathfrak E}_0\cos(\omega t)$.
To this end, we transform into the Kramers-Henneberger frame and calculate 
the coupled Floquet channels numerically.
We find distinct signatures of resonances when the incident energy $E$ 
equals the driving frequency $\omega=E$ which clearly shows the breakdown
of the time-averaged potential approximation. 
As a simple model for experimental applications 
(e.g., in solid state physics), 
we study the rectangular potential, which can also be benchmarked 
with respect to analytical results. 
Finally, we consider the truncated Coulomb potential relevant for nuclear 
fusion.
%
%Keywords: Dynamically Assisted Quantum Tunneling, Assisted Hydrogen Fusion, 
%High-Intensity Electric Fields, Coupled-Channel Equations 
%\\ 
%PACS: {12.20.Ds,11.10.-z, 11.15.Tk}
\end{abstract}

\maketitle
  
%%%%%%%%%%%%%%%%%%%%%%%%%%%%%%%%%%%%%%%%%%%%%%%%%%%%%%%%%%%%%%%%%%%%%%%%%%%%%%%%
%%%%%%%%%%%%%%%%%%%%%%%%%%%%%%%%%%%%%%%%%%%%%%%%%%%%%%%%%%%%%%%%%%%%%%%%%%%%%%%%
\section{Introduction} 
%%%%%%%%%%%%%%%%%%%%%%%%%%%%%%%%%%%%%%%%%%%%%%%%%%%%%%%%%%%%%%%%%%%%%%%%%%%%%%%%
%%%%%%%%%%%%%%%%%%%%%%%%%%%%%%%%%%%%%%%%%%%%%%%%%%%%%%%%%%%%%%%%%%%%%%%%%%%%%%%%

Quantum tunneling is ubiquitous in physics. It plays an important role in many areas, such as  
ultra-cold atoms in optical lattices \cite{Bloch, RevModPhys.78.179,Sias,Ma,Niu,QNS,Eckhardt}, 
electrons in solid-state devices \cite{KaneII, KaneI, ZenerII, Zener, Streetman, Voss, LinderLorke}
as well as nuclear $\alpha$-decay and fusion \cite{G28,GN11,Kelkar,Ivlev+Gudkov}, 
to name only a few examples, see also further applications \cite{doi.org/10.1038/nphys2259, PhysRevLett.107.095301} as well as the reports \cite{doi.org/10.1016/S0370-1573(98)00022-2,Razavy}.
However, even though tunneling through a static potential barrier $V(x)$ 
is usually taught in the basic lecture course on quantum mechanics, 
our understanding of tunneling in time-dependent scenarios such as 
$V(x,t)$ %mostly {\it terra incognita}
is still far from complete.
As one example, let us mention the recent controversy on the dynamical 
assistance of nuclear $\alpha$-decay \cite{PP20, BaiDeng, PhysRevLett.119.202501, Szilvasi, Rizea1, Rizea2}. 
If the temporal variation of the potential $V(x,t)$ is very slow, 
the problem can be treated via the quasi-static potential approximation by 
calculating tunneling as for a static potential barrier
(at each instant of time).
On the other hand, if the potential $V(x,t)$ changes from an initial 
$V_{\rm in}(x)$ to a final form $V_{\rm out}(x)$ in a very short time, 
we may use the sudden approximation by evolving the initial wave function 
$\psi_{\rm in}(x)$ in the final potential $V_{\rm out}(x)$.
Extremely rapidly oscillating potentials $V(x,t)$ can often be dealt within 
the time-averaged potential approximation, where the time-dependent potential 
is replaced by its time average $\overline V(x)=\overline{V(x,t)}$. 

However, away from these extreme cases, one may discover many non-trivial 
and fascinating phenomena \cite{PhysRevD.54.7407, PhysRevLett.67.516}, including resonance \cite{doi.org/10.1006/aphy.2002.6281} or assistance effects \cite{PhysRevE.50.145}.
Furthermore, the regions of validity of the aforementioned approximations 
is also a non-trivial question.
In this respect, the Landauer-B\"uttiker ``traversal'' time \cite{BL82}
plays an important role in separating slow (i.e., adiabatic) 
from fast (i.e., non-adiabatic) processes.

In the following, we study the possible enhancement of tunneling 
by an additional time-dependence, referred to as ``dynamically
assisted tunneling.''
Instead of a fully general space-time-dependent potential $V(x,t)$, 
we consider a static barrier $V(x)$ with an additional 
harmonically oscillating electric field 
${\mathfrak E}(t)={\mathfrak E}_0\cos(\omega t)$.

Note that we restrict our considerations to the directly measurable 
tunneling probability and its increase -- 
but we do not address the question of the ``tunneling time'',
i.e., how long the particle actually stays inside the barrier \cite{BL82, doi.org/10.1038/s41586-020-2490-7, doi.org/10.1016/j.physrep.2006.09.002, doi.org/10.1119/1.1810153, PhysRevLett.119.023201, doi.org/10.1364/OPTICA.1.000343,PhysRevA.107.042216}.
This intriguing question is sometimes also discussed in this context 
and turns out to be quite non-trivial -- already at the level of a 
proper definition. % of the ``tunneling time''. 

To study the enhancement of tunneling, we transform into the 
Kramers-Henneberger frame and employ Floquet theory to split 
the system into a set of coupled channels, see Sec.~\ref{Sec:Form}. 
As our first application, we consider the rectangular potential 
in Sec.~\ref{Sec:Rect} which serves as a simple model for experimental 
applications in solid-state physics, such as scanning tunneling microscopes 
(STM) or tunneling through insulating layers.
As our second application, we study the truncated Coulomb potential 
in Sec.~\ref{Sec:Coulomb} which is relevant for nuclear fusion. 
However, as tunneling phenomena play an important role in many areas,
our main results should also be applicable in other cases, see also 
Secs.~\ref{Sec:FApp} and~\ref{Conclusion}. 

%%%%%%%%%%%%%%%%%%%%%%%%%%%%%%%%%%%%%%%%%%%%%%%%%%%%%%%%%%%%%%%%%%%%%%%%%%%%%%%%
%%%%%%%%%%%%%%%%%%%%%%%%%%%%%%%%%%%%%%%%%%%%%%%%%%%%%%%%%%%%%%%%%%%%%%%%%%%%%%%%
\section{Theoretical Formalism} 
\label{Sec:Form}
%%%%%%%%%%%%%%%%%%%%%%%%%%%%%%%%%%%%%%%%%%%%%%%%%%%%%%%%%%%%%%%%%%%%%%%%%%%%%%%%
%%%%%%%%%%%%%%%%%%%%%%%%%%%%%%%%%%%%%%%%%%%%%%%%%%%%%%%%%%%%%%%%%%%%%%%%%%%%%%%%

Our aim in this paper is to study time-dependent, non-relativistic assisted 
quantum tunneling. 
As such, the  fundamental dynamics of our quantum system is very well governed 
by the Schrödinger equation
\begin{equation}
 \I \hbar\frac{\partial}{\partial t} \psi(x,t) = \hat H(x,t) \psi(x,t),
 %\I \hbar\frac{\partial}{\partial t} \psi(x,t) = \left( - \frac{1}{2m} \frac{\partial^2}{\partial_x^2} + V(x) \right) \psi(x,t),
\end{equation}
with Schrödinger Hamiltonian $\hat H$ and wave function $\psi$. 
For the purposes of this research, we restrict ourselves to a 
single particle (without internal structure) in a
one-dimensional spatial domain in order to limit the rich configuration space. 

We want to further assume that our test particle %is point-like and 
has a defined energy $E$ in its initial state. 
We choose this particular approach as it gives us a clear picture 
of the tunneling probability for a specific energy and provides us 
with simple expressions regarding general scaling laws. 
%Both aspects would be blurred if one were to fixate on, 
%e.g., Gaussian wave packets without any mode decomposition. 
Additionally, a focus on individual energy modes makes it easier to 
relate our findings to discussions on simpler systems, 
such as the well-known tunneling probability 
%cf. the amplitude for transmission 
in static problems \cite{Coleman}
\begin{equation}
 P \sim\mathrm{exp}\left[-2\int {\rm d}x\,\sqrt{2 \mu [V(x)-E]}/\hbar\right],
 \label{eq:coleman}
\end{equation}
with energy $E$, mass $\mu$ and barrier $V(x)$. 
For complementary discussions of assisted tunneling with respect to 
wave packets, we refer to articles~\cite{Grossmann, Longhi, Kohlfurst:2021dfk}. 

Another advantage of focusing on individual modes is that expanding 
the formalism from static to %, in particular, 
time-periodic problems is attainable with relatively little effort, 
see also Refs.~\cite{doi.org/10.1016/j.pquantelec.2004.12.002, doi.org/10.1119/1.15695}. 

In this article we discuss results for a subclass of time-dependent 
potentials \cite{doi.org/10.1063/1.1783592}. 
In particular, we are interested in quantum tunneling in the presence 
of a linearly polarized, oscillating field as described by the
vector potential
\begin{equation}
 A(t) = -\frac{{\mathfrak E}_0}{\omega} \, \sin(\omega t).% \quad \text{with} \quad E(t) = {\mathfrak E}_0 \cos(\omega t) \,.
 \label{eq:A}
\end{equation}
As in many other cases, it is advantageous to study such a 
%Such 
periodically driven quantum system %
%are best studied 
in Floquet theory 
\cite{doi.org/10.1016/S0370-1573(98)00022-2, PhysRevA.56.4045}, 
which is a means of transforming an explicit time-dependent problem 
into a time-independent one. 
The cost comes in the form of an extended Hilbert space as multiple 
energy modes have to be considered within one calculation.

In this case energy is not conserved any more and by means of 
minimal coupling, $\hat p \to \hat p - qA(t)$ with charge $q$, 
a sinusoidal field directly leads to discrete mode coupling, 
i.e., interaction between the Floquet channels.
More precisely, we see the formation of sidebands of energy 
$E \pm n \omega$ with $n$ integer, see Refs.~\cite{Queisser:2019nuh, doi.org/10.1016/S0370-1573(98)00022-2}
or for the related Franz-Keldysh effect \cite{Franz,Keldysh}.

For a static scalar potential $V(x)$ but time-dependent vector potential 
$A(t)$, the Hamiltonian reads ($\hbar=c=1$)
\begin{equation}
\label{original-Hamiltonian}
\hat H(t)=\frac{[\hat p - qA(t)]^2}{2\mu}+V(\hat x)
\,.
\end{equation}
However, in the following we shall work in a different representation.
Similar to the locally freely falling frame -- in which the gravitational 
acceleration is transformed away -- we go to the Kramers-Henneberger frame
where the vector potential $A(t)$ is transformed away.
The price to pay is a time-dependence of the new potential $V(x,t)$.

The basic idea is that, on the basis of the laboratory frame, 
we make an ansatz for the wave function \cite{PhysRevLett.21.838}
\begin{multline}
\psi(x,t) = \exp \bigg(-\frac{\I q^2}{2\mu} \int_{0}^{t} {\rm d} t' A^2(t')\bigg) \\  \times \exp \bigg(\frac{q}{\mu} \int_{0}^{t} {\rm d}t' A(t') \partial_x \bigg)~ \psi_{\rm KH}(x,t),
\end{multline}
where $\psi_{\rm KH}$ is the wave function in the transformed system.
Then, instead of having the wave function being affected by the translation 
operator in the second line,
we can choose the coordinate system that automatically follows 
the new wave function, with the result that the translation operator then 
applies to the potential. 
This removes any explicit reference to the vector potential $A(t)$ 
and converts any field-induced motion into the outcome of a moving 
scalar barrier $V(x,t)$. 
Consequently, the Schrödinger equation reads
\begin{multline}
\I \hbar \frac{\partial}{\partial t} \psi_{\rm KH}(x,t) 
= \\ 
\bigg(-\frac{1}{2\mu} \frac{\partial^2}{\partial_x^2} 
+ V \big[x - \chi(t) \big] \bigg) \psi_{\rm KH}(x,t)
\,,
\end{multline}
where the displacement $\chi(t)$ satisfies the classical equation 
of motion $\mu\ddot\chi(t) = q\dot A(t)$
for a point-particle in a time-dependent external field.

The mathematics behind this change in reference frame is provided in 
Appendix~\ref{Sec:KH}.
Further note the difference between a potential moving in spatial 
direction and a scalar potential of modulated height 
\cite{Pimpale, PhysRevB.26.6408, PhysRevB.60.15732}.

Working in the Kramers-Henneberger frame has several advantages. 
On the one hand, it is more intuitive as we only have to account 
for a single rigid potential with a time-dependent displacement.
%moving over time. 
%
On the other hand, for spatially localized potential barriers, 
the effect of the time-dependent electric field is also spatially localized.
In contrast, the time-dependence of the original 
Hamiltonian~\eqref{original-Hamiltonian} applies to all space points 
in the same way.
Even worse, the representation of the electric field ${\mathfrak E}(t)$
via the potential ${\cal V}(x,t) = x \, q{\mathfrak E}(t)$ would result in 
a ``perturbation'' which increases without bound at spatial infinity.

At this point we apply Floquet theory in order to factor out the time 
evolution and reduce the Schrödinger equation to a system of ordinary 
differential equations. 
To do so, we expand the wave function into an unperturbed part times 
a Fourier decomposition 
\begin{equation}
\psi_{\rm KH}(x,t) = \ee^{-\I Et} \sum_{n=-\infty}^{\infty} u_n(x) \ee^{\I n\omega t},
\label{eq:psiKH}
\end{equation}
and, following the same principle, we also decompose the new scalar 
potential into Fourier modes
\begin{equation}
V(x,t) = \sum_{n=-\infty}^{\infty} V_n (x)\,\ee^{\I n\omega t}.
\end{equation}
The resulting equations do not exhibit an explicit time-dependence any more 
\begin{equation}
\frac{{\rm d}^2}{{\rm d} x^2} u_n(x) + k_{n}^{2} u_n(x) = \sum_{m=-\infty}^{\infty} v_{nm}(x) u_m(x),
\label{eq:u}
\end{equation}
with $k_n \equiv \sqrt{2 \mu (E+n\omega)}$ and 
$v_{nm}(x) \equiv 2 \mu \,V_{n-m}(x)$. 
Thus, %in the Floquet approach 
instead of solving a partial differential 
equation in $x$ and $t$ we solve a system of mode-coupled, 
ordinary differential equations in $x$ only. 

We proceed by re-writing this equation from a boundary-valued 
to a new initial-valued problem. 
For the sake of simplicity, we will assume that we integrate 
over a sufficiently large spatial interval such that any given 
scalar potentials have either become zero or have fallen off 
substantially at the boundaries of integration. 
In this case, we can assume that mode functions become decoupled 
plane waves asymptotically. 

Ultimately, our goal is to find an expression for the particle's 
tunneling probability and not necessarily the wave function itself. 
Additionally, to automatically take into account boundary conditions 
(such that there are only incoming waves from the left), 
we reformulate Eq.~\eqref{eq:u} as an integro-differential equation 
\begin{multline}
u_{nl}(x) = \ee^{\I k_n x} \delta_{nl} \\
+
%\frac{1}{2 \I k_n}
\sum_{m=-\infty}^{\infty} \int_{-\infty}^{\infty} {\rm d}x'
\  G(x,x')
%\ee^{\I k_n \lvert x-x' \rvert} 
v_{nm}(x') u_{ml}(x')
\,,
\label{eq:uni}
\end{multline}
in terms of the Green's function
$G(x,x')=\ee^{\I k_n \lvert x-x' \rvert}/(2 \I k_n) $
of the differential operator in Eq.~\eqref{eq:u}.

Instead of solving Eq.~\eqref{eq:uni} directly, we introduce 
an auxiliary 
variable $y$ with the intention 
to progressively probe the full potential
\begin{multline}
u_{nl}(x,y) = \ee^{\I k_n x} \delta_{nl} \\
+ \frac{1}{2 \I k_n} \sum_{m=-\infty}^{\infty} \int_{y}^{\infty} {\rm d}x'
\ \ee^{\I k_n \lvert x-x' \rvert} v_{nm}(x') u_{ml}(x',y)
\,.
\label{eq:aux}
\end{multline}
Taking $y$-derivatives and using the fact that we are only interested in 
the asymptotic limits $x\to\pm\infty$, we may map the integro-differential equations~\eqref{eq:uni} onto a set of ordinary differential equations in $y$ in terms of $y$-dependent quasi-transmission and quasi-reflection amplitudes (see also Ref.~\cite{Razavy} and Appendix~\ref{Sec:Framework})
\begin{align}
\begin{split}
\frac{{\rm d}}{{\rm d}y} \rho_{nl}(y) {}&= 
 - \sum_{s=-\infty}^{\infty} \frac{1}{2 \I k_s} \bigg(e^{\I k_s y}\delta_{ns} + \rho_{ns}(y) e^{-\I k_s y}  \bigg) \\
  & \hspace{-0.3cm} \times \sum_{m=-\infty}^{\infty} v_{sm}(y)\bigg(e^{\I k_m y}\delta_{ml} + \rho_{ml} (y)e^{-\I k_m y}  \bigg),
  \label{eq:r}
  \end{split}\\
 \begin{split}  
\frac{d}{dy} \tau_{nl}(y) {}&= 
- \sum_{s=-\infty}^{\infty} \frac{1}{2 \I k_s} \bigg(
e^{-\I k_s y}\delta_{ns}+\tau_{ns}(y) e^{-\I k_s y} \bigg)\\
  & \hspace{-0.3cm} \times \sum_{m=-\infty}^{\infty} v_{sm}(y) \bigg(e^{\I k_m y}\delta_{ml} + \rho_{ml}(y)e^{-\I k_m y}  \bigg).
  \label{eq:t}
\end{split}
\end{align}
This coupled set of equations is now more suitable for numerical integration. 
According to Eq.~\eqref{eq:aux}, we are interested in the behavior at 
$y \to -\infty$ while the other limit $y \to \infty$ becomes trivial, 
leading to the initial conditions 
$\rho_{nl}(y \to \infty) = 0$ and $\tau_{nl}(y \to \infty) = 0$. 
Thus, we have to integrate Eqs.~\eqref{eq:r} and~\eqref{eq:t} 
backwards from $y \to \infty$ to $y \to -\infty$.

In the asymptotic limit we recover the transmission $T_{nl}$ 
and reflection $R_{nl}$ amplitudes for the various channels
\begin{equation}
\quad T_{nl} = \delta_{nl}+\tau_{nl} (y \to -\infty)
\,,\;
R_{nl} = \rho_{nl} (y \to -\infty)
\,.
\end{equation}
More precisely, the probability for a mode of initial energy $E_l$ 
to become a transmitted or reflected mode with final energy $E_n$ 
are given by $\lvert T_{nl} \rvert^2 k_n/k_l $ or 
$ \lvert R_{nl} \rvert^2 k_n/k_l$, respectively.
It should be noted, however, that this relation is only valid for matter 
waves with real and positive wave vectors $k_n$ and $k_l$.

In the following, we focus on the transmission amplitudes $T_{nl}$.
However, the formalism can also be used to study the 
reflective properties of a potential, see Sec.~\ref{Sec:FApp} below. 
A detailed derivation of the formalism can be found in Appendix~\ref{Sec:Framework}. 
For a more detailed review, see Ref.~\cite{Razavy}. 
Computational solution techniques, error analysis, as well as the 
performance and efficiency of the used solver are discussed in 
Ref.~\cite{Ryndyk}.

%%%%%%%%%%%%%%%%%%%%%%%%%%%%%%%%%%%%%%%%%%%%%%%%%%%%%%%%%%%%%%%%%%%%%%%%%%%%%%%%
%%%%%%%%%%%%%%%%%%%%%%%%%%%%%%%%%%%%%%%%%%%%%%%%%%%%%%%%%%%%%%%%%%%%%%%%%%%%%%%%
\section{Rectangular Potential} 
\label{Sec:Rect}
%%%%%%%%%%%%%%%%%%%%%%%%%%%%%%%%%%%%%%%%%%%%%%%%%%%%%%%%%%%%%%%%%%%%%%%%%%%%%%%%
%%%%%%%%%%%%%%%%%%%%%%%%%%%%%%%%%%%%%%%%%%%%%%%%%%%%%%%%%%%%%%%%%%%%%%%%%%%%%%%%

We first consider scattering at a rectangular potential
\begin{equation}
V(x) = 
 \begin{cases} 
  0 & \text{for } x \leq 0, \\ 
  V_0>0 & \text{for } 0 < x < L, \\ 
  -V_1 & \text{for } x \geq L,  
 \end{cases}
 \label{eq:rect}
 \end{equation}
of height $V_0$ and length $L$. 
The asymptotic potential levels before and after the barrier can be different,
given by the off-set $V_1$.
This potential~\eqref{eq:rect} can be considered as a one-dimensional 
toy model for a scanning 
tunneling microscope (STM) where the region $x\leq0$ corresponds to the tip and  
the region $x\geq L$ to the substrate, while the interval $0 < x < L$ models 
the vacuum in between.
As another application, the potential could describe tunneling
through an insulating layer between two conductors.

As a standard textbook example of quantum tunneling, the rectangular potential 
is particularly interesting as a toy model. 
In the static case, the exact reflection and transmission coefficients 
are well known. 
In the dynamical case of assisted tunneling, especially in the 
Kramers-Henneberger framework the physical implications of having an 
additional external field present can be understood intuitively. 
The lack of an internal structure in $V_0$ and the clear separation between 
regions (before, after and inside the barrier) is a feature that is quite 
beneficial for computational techniques and even facilitates analytical
results in various limiting cases, see Appendices~\ref{Sec:Rec0}-\ref{Sec:Rec3}.

Before discussing the results for the rectangular potential, let us briefly 
recapitulate how an electric field ${\mathfrak E}(t)$ could enhance (or suppress)
tunneling. 
Trying to separate the different effects, one can identify at least four main 
contributions, see also Ref. \cite{Kohlfurst:2021dfk}. 
Even if we ignore the time-dependence of the electric field ${\mathfrak E}$,
it can accelerate (or decelerate) the particle before it hits the barrier 
({\it pre-acceleration}) and thereby enhance (or suppress) tunneling, 
or it can effectively deform the barrier ({\it deformation}).
Apart from these two adiabatic effects, the time-dependence of ${\mathfrak E}(t)$
induces a coupling of the Floquet channels, leading to an effectively higher 
(or lower) energy $E\to E \pm n\omega$.
This {\it energy mixing} is most effective at the front end of the barrier.
In addition, the Kramers-Henneberger {\it displacement} is most effective at 
the rear end of the barrier, where it can liberate the wave packet inside the 
barrier and thereby enhance tunneling.

%%%%%%%%%%%%%%%%%%%%%%%%%%%%%%%%%%%%%%%%%%%%%%%%%%%%%%%%%%%%%%%%%%%%%%%%%%%%%%%%
\subsection{Resonance Effects} 
\label{Sec:Reso}
%%%%%%%%%%%%%%%%%%%%%%%%%%%%%%%%%%%%%%%%%%%%%%%%%%%%%%%%%%%%%%%%%%%%%%%%%%%%%%%%

\begin{figure}[t]
 \includegraphics[trim={0cm 0.0cm 0cm 0.cm},clip,width = 0.49\textwidth]{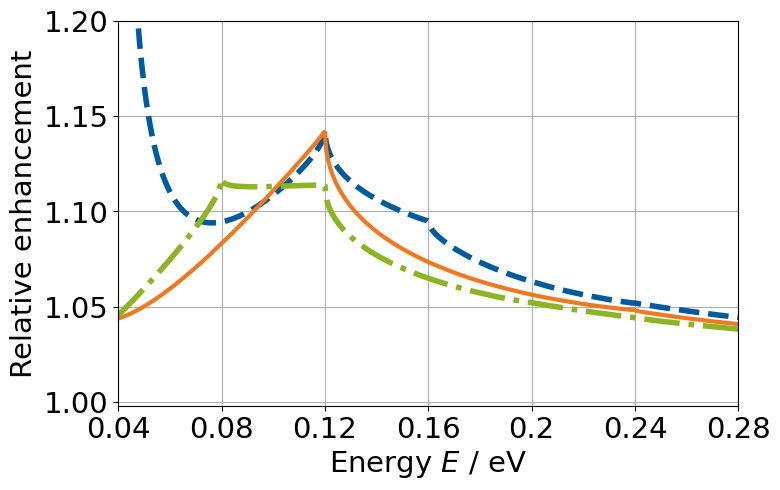}
 \caption{
 Relative enhancement in the electron transmission probability through a rectangular barrier
 as a function of initial energy $E$ for different asymptotic levels $V_1$. 
 The blue, dashed curve corresponds to tunneling through a barrier where the 
 potential behind the barrier is higher than the initial potential $V_1<0$. 
 The green dot-dashed curve shows tunneling through a barrier with a 
 lower-laying asymptotic potential $V_1>0$. 
 The orange, solid line holds as a reference where asymptotic levels are 
 equal $V_1=0$. 
 One resonance appears where the incident energy $E$ of the matter wave 
 and the driving frequency $\omega$ coincide. 
 The second peak appears around energies of $E=\omega-V_1$. 
 %For $V_1<0$, one can also observe another peak at $E=-V_1$. 
 Parameters: barrier height $V_0 = 6$ eV, barrier width $L = 0.2$ nm and field strength $\mathfrak{E}_0 = 2.4 \times 10^8$ V/m.
 }
 \label{fig:B5}
\end{figure}

Let us now present the numerical results for the rectangular potential,
obtained by solving Eqs.~\eqref{eq:r} and~\eqref{eq:t} for the coupled Floquet 
channels.
The relative enhancement, i.e., dynamically assisted transmission probability
divided by the value without the field ${\mathfrak E}(t)$, is plotted in 
Fig.~\ref{fig:B5} for a specific barrier size and field strength and three values for the off-set $V_1$.

%%%%%%%%%%%%%%%%%%%%%%%%%%%%%%%%%%%%%%%%%%%%%%%%%%%%%%%%%%%%%%%%%%%%%%%%%%%%%%%%
%%%%%%%%%%%%%%%%%%%%%%%%%%%%%%%%%%%%%%%%%%%%%%%%%%%%%%%%%%%%%%%%%%%%%%%%%%%%%%%%
 
As a first result, we do see a significant enhancement of the tunneling 
probability for the chosen set of parameters. 
For other parameters (e.g., higher field strengths ${\mathfrak E}_0$), 
the enhancement can even be more pronounced, but this typically requires
taking into account more Floquet channels and is thus numerically more 
demanding.

As a second result, we observe resonance effects in the form of peaks or cusps 
at $E=\omega$ and $E=\omega-V_1$. % (as well as $E=-V_1$).
These features clearly show that the time-averaged potential approximation 
cannot be applied here -- which is perhaps not too surprising since the 
driving frequency $\omega$ is not much larger than all other relevant scales.  
Furthermore, the non-monotonic dependence on $E$ indicates that this feature 
can also not be explained by the adiabatic pre-acceleration and potential 
deformation effects mentioned above. 
The resonance effects are of non-adiabatic origin and occur when a Floquet 
channel opens up or closes in one or both of the asymptotic regions.  
The dependence on $V_1$ shows that energy mixing at the front end is probably 
not the only relevant mechanism here, the displacement (which also occurs at 
the rear end) must also be relevant.
Quite intuitively, the Floquet channel which opens up or closes corresponds 
to a very large wavelength outside the barrier, which makes it very susceptible 
to these non-adiabatic effects. This is true even for evanescent waves near the 
threshold, which can make a significant contribution to the transmission probability 
if the corresponding energy band is populated and the probability is consequently transferred 
to an out-going wave.

\begin{figure}[t]
 \includegraphics[width = 0.49\textwidth]{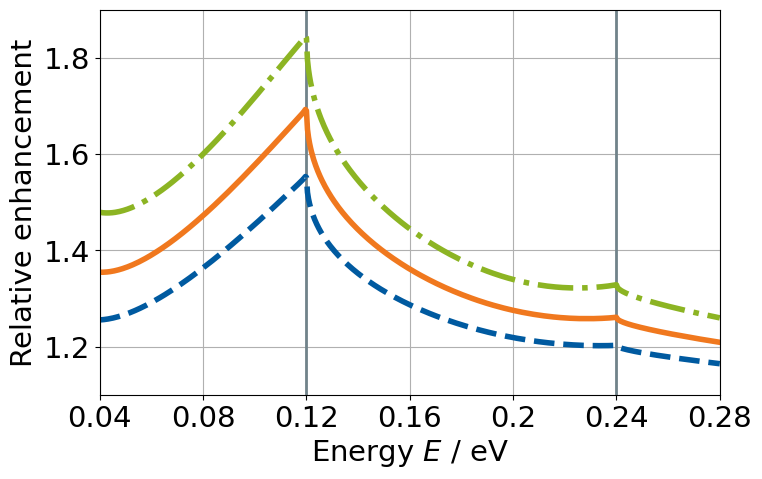}
 \caption{
 Relative enhancement in the electron transmission probability through a rectangular barrier
 as a function of initial energy $E$ with emphasis on higher-order resonance effects. 
 The field strength ${\mathfrak E}_0$ is varied from $\mathfrak{E}_0 = 4.8 \times 10^8$ V/m (blue, dashed) 
 via $\mathfrak{E}_0 = 5.4 \times 10^8$ V/m (orange, solid) to $\mathfrak{E}_0 = 6.0 \times 10^8$ V/m (green, dot-dashed), with field frequency $\omega = 0.12$ eV, 
 barrier length $L = 0.2$ nm 
 and barrier height $V_0 = 6$ eV being fixed. 
 At $V_1=0$ optimal relative enhancement is found at $E = \omega$ as well as 
 $E = 2\omega$, i.e., at $0.12$ eV as well as $0.24$ eV (grey vertical lines).
 }
 \label{fig:B1}
\end{figure}

%\sout{A closer inspection of Fig.~\ref{fig:B5} reveals second-order resonances at $E=2\omega$, but they are very hard to see for this set of parameters.Thus, we plotted another set of parameters in Fig.~\ref{fig:B1} where these second-order resonances are better visible.}
A closer inspection of Fig.~\ref{fig:B5} reveals higher-order resonances at 
$E=2\omega$, but as they are hard to see we plotted another set of parameters 
in Fig.~\ref{fig:B1} with special emphasis on second-order resonances.

%%%%%%%%%%%%%%%%%%%%%%%%%%%%%%%%%%%%%%%%%%%%%%%%%%%%%%%%%%%%%%%%%%%%%%%%%%%%%%%%
%%%%%%%%%%%%%%%%%%%%%%%%%%%%%%%%%%%%%%%%%%%%%%%%%%%%%%%%%%%%%%%%%%%%%%%%%%%%%%%%

%%%%%%%%%%%%%%%%%%%%%%%%%%%%%%%%%%%%%%%%%%%%%%%%%%%%%%%%%%%%%%%%%%%%%%%%%%%%%%%%
\subsection{Side-Bands} 
\label{Sec:Side}
%%%%%%%%%%%%%%%%%%%%%%%%%%%%%%%%%%%%%%%%%%%%%%%%%%%%%%%%%%%%%%%%%%%%%%%%%%%%%%%%

%%%%%%%%%%%%%%%%%%%%%%%%%%%%%%%%%%%%%%%%%%%%%%%%%%%%%%%%%%%%%%%%%%%%%%%%%%%%%%%%
\begin{figure}[t]
 \includegraphics[trim={0cm 0.0cm 0cm 0.cm},clip,width = 0.49\textwidth]{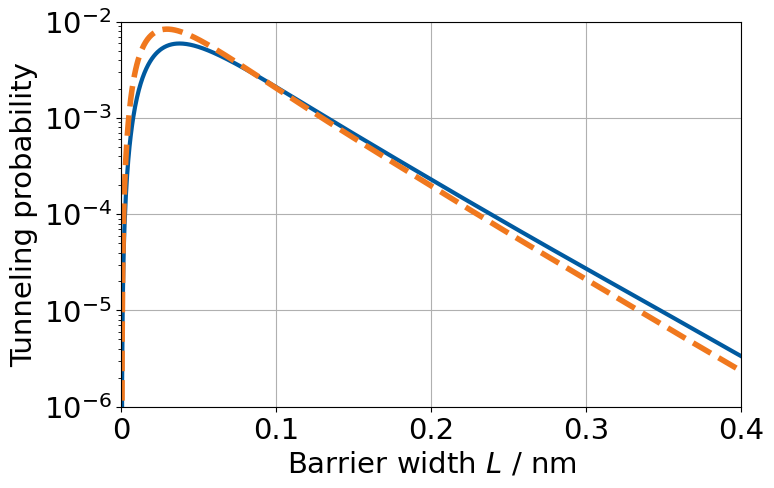} 
 \caption{
 Contribution of the first sidebands on each side, 
 $E + \omega$ (blue, solid) and $E - \omega$ (orange, dashed), 
 to the probability for electron transmission through a rectangular barrier 
 as function of the barrier width $L$. 
 Parameters: barrier height $V_0 = 6$ eV, field frequency $\omega=0.12$ eV, field strength $\mathfrak{E}_0 = 6 \times 10^8$ V/m, energy $E = 0.28$ eV.
 }
 \label{fig:B2} 
\end{figure}
%%%%%%%%%%%%%%%%%%%%%%%%%%%%%%%%%%%%%%%%%%%%%%%%%%%%%%%%%%%%%%%%%%%%%%%%%%%%%%%%

Now, after having discussed the total transmission probability, let us study 
the contributions of the side-bands, where we focus on the first two side-bands 
$E\pm\omega$ for simplicity, as shown in Fig.~\ref{fig:B2} for 
vanishing off-set $V_1=0$. 

The lower side-band at $E-\omega$ 
dominates for short barriers, only for long enough barriers the upper 
side-band at $E+\omega$ takes over. 
To understand the origin for this behavior, let us first compare it to 
analytical approximations in the corresponding limiting cases. 

For small quiver amplitudes $\chi_0$ and short barriers, we may obtain the 
transmission amplitudes for the first side-bands via perturbation theory,
see Appendix~\ref{Sec:Rec2}
\begin{equation}
 T_{\pm 1}^{\rm transparent} \propto 
 \left( L \chi_0 \mu V_0 \right)^2\,\frac{E}{E \pm \omega}
 \,.
\end{equation}
This result confirms that the lower side-band dominates in this regime 
and points to the following intuitive interpretation:
For the lower side-band, the temporal oscillation period and the 
wavelength of the mode (outside the barrier) are longer 
than for the upper side-band and thus the lower side-band is more 
susceptible to the Kramers-Henneberger displacement $\chi(t)$ and 
responds stronger than the upper side-band. 

On the other hand, the opaque-barrier limit is reached for long barriers 
of sufficient height $V_0\gg\omega,E$.
In this limit, we obtain the approximations, see Appendix~\ref{Sec:Rec3}
\begin{align}
 \label{eq:upper}
 & T_{+1}^{\rm opaque} \propto 
 \frac{\chi_0 E}{V_0} (V_0 - E - \omega) 
 \ee^{-2 \sqrt{2 \mu(V_0 - E - \omega)} L },\\
 & T_{-1}^{\rm opaque} \propto 
 \frac{\chi_0 E}{V_0} (V_0 - E) 
 \ee^{-2 \sqrt{2 \mu(V_0 - E)} L}.
 \label{eq:lower} 
\end{align}
In search for an intuitive explanation, we may understand the 
exponent~\eqref{eq:upper} in the following way: 
At the front end of the barrier, the incident wave with energy $E$ 
is up-shifted to $E+\omega$ by the oscillating Kramers-Henneberger
displacement (energy mixing) and thus can tunnel easier through 
the barrier, see Eq.~\eqref{eq:coleman}. 
Applying the same argument to the lower side-band~\eqref{eq:lower},
one might expect an exponent 
$\ee^{-2 \sqrt{2 \mu(V_0 - E + \omega)} L }$. 
However, this is not the dominant contribution in this case. 
Instead, it stems from the central band with energy $E$ which 
tunnels through the barrier 
with the usual exponent $\ee^{-2 \sqrt{2 \mu(V_0 - E)} L }$ and is 
down-shifted to $E-\omega$ at the rear end of the barrier.

In conclusion, the various bands display different dependences on the parameters 
and the upper bands are not always favored over the lower energy modes. 
Nevertheless, these first results already show that by pumping energy into 
the system, the overall probability of tunneling is typically 
increased in comparison to static tunneling.

%%%%%%%%%%%%%%%%%%%%%%%%%%%%%%%%%%%%%%%%%%%%%%%%%%%%%%%%%%%%%%%%%%%%%%%%%%%%%%%%
%%%%%%%%%%%%%%%%%%%%%%%%%%%%%%%%%%%%%%%%%%%%%%%%%%%%%%%%%%%%%%%%%%%%%%%%%%%%%%%%
\section{Assisted Nuclear Fusion} 
\label{Sec:Coulomb}
%%%%%%%%%%%%%%%%%%%%%%%%%%%%%%%%%%%%%%%%%%%%%%%%%%%%%%%%%%%%%%%%%%%%%%%%%%%%%%%%
%%%%%%%%%%%%%%%%%%%%%%%%%%%%%%%%%%%%%%%%%%%%%%%%%%%%%%%%%%%%%%%%%%%%%%%%%%%%%%%%

Another potential application of strong-field induced assisted tunneling is in 
nuclear fusion \cite{Queisser:2019nuh, Kohlfurst:2021dfk, PhysRevC.105.054001, doi.org/10.1038/s41598-022-08433-4, PhysRevC.106.034003, PhysRevC.102.011601}.
Since the characteristic length scale of tunneling is very short 
(in the sub-pm regime), we neglect the impact of the surrounding particles 
and treat nuclear fusion as a two-particle process. \cite{Laser1}
Hence, the probability for, e.g., deuterium (D) and tritium (T) to merge can, 
at low temperatures, be modeled as non-relativistic Schrödinger tunneling 
through the Coulomb barrier between the two positively charged particles \cite{Guire, Giuffrida, Moreau_1977, Hora1, Labaune}.  

Specifically, we model assisted DT-fusion through non-relativistic 1D  
quantum mechanics where initially we have two point-particles with masses 
$m_{\rm D}$ and $m_{\rm T}$ at positions $r_{\rm D}$ and $r_{\rm T}$ 
with velocities $\dot{r}_{\rm D}$ and $\dot{r}_{\rm T}$, respectively. 
In this case, particle dynamics are given by a two-particle Lagrangian
\begin{multline}
 \mathcal L_{DT} = 
  \frac{m_{\rm D} \dot{r}_{\rm D}^2}{2}\,
  +
  \frac{m_{\rm T}\dot{r}_{\rm T}^2}{2}\, \\
  -
  V(\lvert r_{\rm D}-r_{\rm T} \rvert)
  +q
  \dot{r}_{\rm D} A(t,r_{\rm D})
  +
  q\dot{r}_{\rm T} A(t,r_{\rm T}) ,
\end{multline}
with kinetic terms in the first line and couplings 
to the potentials in the second. 
Hereby, $V(\lvert r_{\rm D}-r_{\rm T} \rvert)$ contains the Coulomb 
repulsion as well as nuclear attraction. 
The vector potential $A$ corresponds, at this point, to a general 
electromagnetic field with a field strength shortly below the critical 
field strength of $E_{\rm cr} \sim 1.3 \times 10^{18}$~V/m and a 
characteristic frequency in the keV-regime. \cite{Laser2}

%%%%%%%%%%%%%%%%%%%%%%%%%%%%%%%%%%%%%%%%%%%%%%%%%%%%%%%%%%%%%%%%%%%%%%%%%%%%%%%%
\begin{figure}[t]
 \includegraphics[width = 0.49\textwidth]{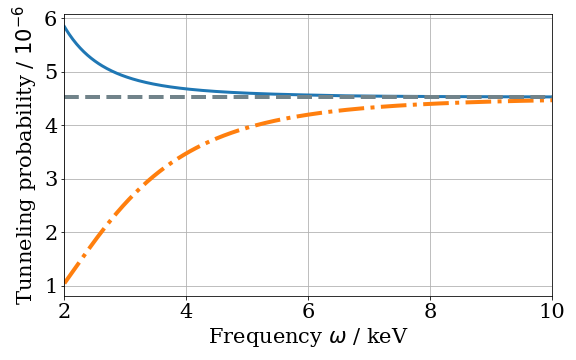}
 \caption{
 Total transmission probability in a truncated Coulomb 
 potential~\eqref{eq:truncated} with $V_1=0$ 
 as a function of quiver frequency $\omega$ comparing the full numerical 
 simulation (blue, solid line) with the static field-free solution 
 (straight, grey, dashed line) and the time-averaged potential approximation
 (orange, dot dashed line). 
 The latter refers to a static potential where the quiver dynamics 
 has been averaged over, which effectively amounts to keeping the zeroth 
 Floquet channel only and neglecting all the others.
 %integrated out, thus only one channel $N=1$ remains. 
 For the blue curve a total of $N=33$ interacting energy channels have been used.
 Initial energy $E$ is $6$ keV and the field strength is 
 ${\mathfrak E}_0 \approx 2 \times 10^{16}$~V/m.}
 \label{fig:AppAC4}
\end{figure}
%%%%%%%%%%%%%%%%%%%%%%%%%%%%%%%%%%%%%%%%%%%%%%%%%%%%%%%%%%%%%%%%%%%%%%%%%%%%%%%%

To simplify computation we introduce center-of-mass 
$R=(m_{\rm D}r_{\rm D}+m_{\rm T}r_{\rm T})/(m_{\rm D}+m_{\rm T})$ 
and relative coordinates $r=r_{\rm D}-r_{\rm T}$, 
respectively. 
Since the characteristic distances $r$ in the sub-pm regime are
much smaller than the wavelength of an x-ray field $A$ with 
frequencies of order keV, we may approximate $A(t,r_{\rm D})\approx A(t,R)$
and $A(t,r_{\rm T})\approx A(t,R)$. 
As a result, the center-of-mass motion completely decouples from the relative 
coordinate~$r$.
The dynamics of assisted fusion is thus entirely described 
by an effective one-particle Lagrangian
\begin{equation}
\mathcal L
=
\frac{\mu}{2}\,\dot{r}^2
-
V(\lvert r \rvert )
+
q_{\rm eff} \dot{r} A(t)
,
\end{equation}
with reduced mass 
$\mu=(m_{\rm D}^{-1}+m_{\rm T}^{-1})^{-1} \approx 1.13~\si{\giga\electronvolt}$ 
and effective charge 
$q_{\rm eff}=q(m_{\rm T}-m_{\rm D})/(m_{\rm T}+m_{\rm D})\approx q/5$,
while $A(t)=A(t,R)$ is effectively purely time-dependent.

The full potential is given by
\begin{equation}
\label{eq:truncated}
V(r) = 
 \begin{cases} 
  \mbox{$\displaystyle\frac{q^2}{4\pi \varepsilon_0}\,\frac{1}{r}$} 
  %1/(4\pi \varepsilon_0) \, q^2 / r 
  & \text{for } r \geq r_0, \\ 
  -V_1 & \text{for } r<r_0,  
 \end{cases}
\end{equation}
with the inner turning point $r_0 = 3.89$~fm representing the region
where nuclear attraction, modeled here by a constant potential $V_1$, 
takes over \cite{PhysRev.94.737}. 
For a DT-system the peak of the Coulomb potential at $r_0$
is roughly $375$~keV. 
This is much larger than both the particle energies as well as the field 
frequencies which are of the order of $\mathcal O (1-10)$~keV. 
It is much lower, however, when compared to the depth of the nuclear potential 
which is in the MeV-regime. 

%%%%%%%%%%%%%%%%%%%%%%%%%%%%%%%%%%%%%%%%%%%%%%%%%%%%%%%%%%%%%%%%%%%%%%%%%%%%%%%%
\begin{figure}[t]
 \includegraphics[width = 0.49\textwidth]{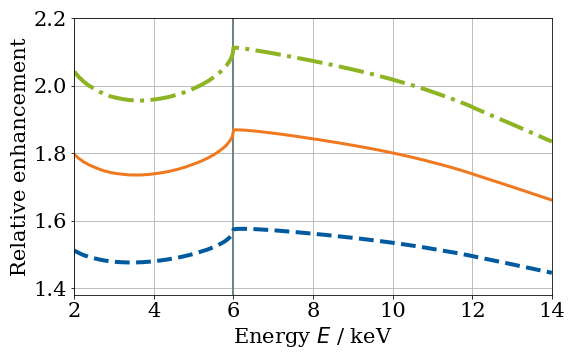}
 \caption{
 Relative enhancement in the particle transmission probability 
 in a truncated Coulomb potential with $V_1=0$ as a function 
 of initial energy $E$ for $\omega=6$~keV and $V_1=0$, plotted for  
 three different field strengths 
 ${\mathfrak E}_0$ (green, dot-dashed: $2.0\times 10^{17}$~V/m; 
 orange, solid: $1.8\times 10^{17}$~V/m; 
 blue, dashed: $1.5\times 10^{17}$~V/m). 
 Again, at $E=\omega$ (vertical, grey line) resonance peaks are found.
 Sufficiently away from the peaks, the relative enhancement of the transmission 
 probability decreases with energy $E$.
 }
 \label{fig:AC2}
\end{figure}
%%%%%%%%%%%%%%%%%%%%%%%%%%%%%%%%%%%%%%%%%%%%%%%%%%%%%%%%%%%%%%%%%%%%%%%%%%%%%%%%

Even though analytical studies of the Coulomb potential are more involved
than for the rectangular potential, the reflection and transmission amplitudes 
for unassisted tunneling can be obtained analytically in terms of Whittaker 
functions, cf.~Appendix~\ref{Sec:AppTrunc}. 
However, we found a direct numerical integration of the set of differential 
equations \eqref{eq:r}-\eqref{eq:t} to be faster and more accurate than 
evaluating the Whittaker functions.

%%%%%%%%%%%%%%%%%%%%%%%%%%%%%%%%%%%%%%%%%%%%%%%%%%%%%%%%%%%%%%%%%%%%%%%%%%%%%%%%
\subsection{Comparison to Time-averaged Potential} 
\label{Sec:Time}
%%%%%%%%%%%%%%%%%%%%%%%%%%%%%%%%%%%%%%%%%%%%%%%%%%%%%%%%%%%%%%%%%%%%%%%%%%%%%%%%

In Fig.~\ref{fig:AppAC4} we display an exemplary result on assisted 
nuclear fusion. 
We want to place emphasis specifically on the different behavior when 
taking into account the full set of coupled differential equations in 
comparison to the time-averaged potential approximation, 
%only considering an averaged potential, 
or equivalently, only considering
the null mode in a Fourier decomposition of the Kramers-Henneberger 
Coulomb potential, see also the discussion in Refs.~\cite{PhysRevC.100.064610, Queisser:2020mns}. 
The latter is an approximation that has been widely used, 
for example in laser-induced atomic stabilization, 
because of the enormous computational simplifications it brings. 
For an overview, see the articles~\cite{PhysRevLett.66.1038, Eberly} 
or the review~\cite{Gavrila} and references therein. 
However, for tunneling through a laterally moving barrier, 
as is the case in this article, Fig.~\ref{fig:AppAC4} clearly shows that such a 
technical simplification leads to an incorrect tunneling probability 
in the low-frequency regime. 
The simple reason is that considering only the zero mode results in an 
effectively higher tunneling potential, which in turn results in a 
lower overall transmission probability. 
This is the opposite of what is observed in a complete simulation 
where sidebands and the interaction of the main channel with the 
sidebands result in a net increase in transmission.

Note that the failure of the time-averaged potential approximation
at low frequencies is not too surprising since this approximation is 
valid when the oscillation frequency $\omega$ is much faster than all 
other relevant energy or frequency scales. 

As another point, the result in Fig.~\ref{fig:AppAC4} was obtained for vanishing 
off-set $V_1=0$.
Introducing such an off-set $V_1>0$, the time-averaged potential approximation 
could result in a tunneling probability which is higher than in the static case 
because the time-averaged potential is then reduced near its maximum. 
However, in this situation the applicability of the time-averaged potential 
approximation is even more questionable:
For a reduced mass of approximately 1~GeV, a potential drop $V_1$ of 10~MeV 
or more implies that the wave packet moves away from the barrier 
(after tunneling) with a large velocity of around one seventh of the 
speed of light or more.
This is faster than the velocity of the quiver motion
induced by the XFEL (with a frequency of 2~keV or more), 
unless field strengths above the Schwinger limit 
$E_{\rm cr} \sim 1.3 \times 10^{18}$~V/m are considered. 
Thus, the wave packet has no chance to experience a time-averaged
potential.

%%%%%%%%%%%%%%%%%%%%%%%%%%%%%%%%%%%%%%%%%%%%%%%%%%%%%%%%%%%%%%%%%%%%%%%%%%%%%%%%
\subsection{Resonance Effects} 
\label{Sec:Res-nucl}
%%%%%%%%%%%%%%%%%%%%%%%%%%%%%%%%%%%%%%%%%%%%%%%%%%%%%%%%%%%%%%%%%%%%%%%%%%%%%%%%

As we have discussed in Sec.~\ref{Sec:Reso}, resonance effects can occur 
at energies $E=n\omega$ where Floquet channels open up (or close).
This phenomenon is not tied to the rectangular potential but can also be 
observed in assisted nuclear fusion, cf.~Fig.~\ref{fig:AC2} 
for the case of vanishing off-set $V_1=0$.
One has to be careful in interpreting these results, however, as the Coulomb 
potential is very asymmetric. 
Consequently, non-adiabatic enhancement effects caused by the front end 
of the barrier are suppressed due to the slow and gradual change of the 
potential at the outer turning point $r^\ast=q^2/(4\pi\varepsilon_0E)$, 
see also Ref.~\cite{Kohlfurst:2021dfk}. 
In contrast, the sharp drop of the barrier at its rear end, 
i.e., the inner turning point $r_0$, facilitates strong non-adiabatic 
enhancement effects. 
As an intuitive picture, the Kramers-Henneberger motion $\chi(t)$ of the 
barrier may ``push'' part of the wave function out of the rear of the barrier
(displacement effect).

A technical discussion on the eigenvalue structure of the corresponding 
Schrödinger equation and their connection to resonances is further discussed 
in Appendix~\ref{Sec:Resonances}. 

%%%%%%%%%%%%%%%%%%%%%%%%%%%%%%%%%%%%%%%%%%%%%%%%%%%%%%%%%%%%%%%%%%%%%%%%%%%%%%%%
\subsection{Influence of Off-set $V_1$} 
\label{Sec:off-set} 
%%%%%%%%%%%%%%%%%%%%%%%%%%%%%%%%%%%%%%%%%%%%%%%%%%%%%%%%%%%%%%%%%%%%%%%%%%%%%%%%

%%%%%%%%%%%%%%%%%%%%%%%%%%%%%%%%%%%%%%%%%%%%%%%%%%%%%%%%%%%%%%%%%%%%%%%%%%%%%%%%
\begin{figure}[t]
 \includegraphics[width=0.49\textwidth]{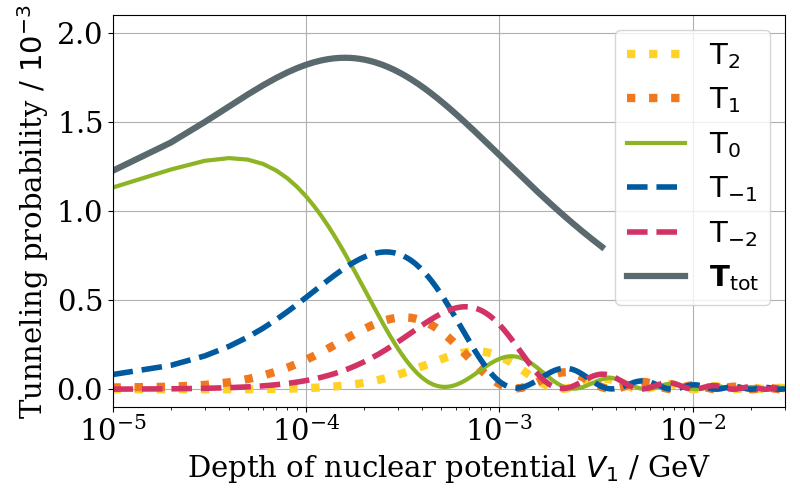}
 \caption{
 Transmission probability in assisted DT-fusion as a function of the depth of 
 the attractive nuclear potential $V_1$. 
 Important individual channels, indicated by their difference from the initial 
 energy in terms of $\pm n\omega$, as well as the total transmission 
 probability (grey, solid line) are shown. 
 The initial particle energy is $E=14$~keV, while the applied electric field 
 shows a peak field strength of ${\mathfrak E}_0 \approx 8\times 10^{16}$~V/m 
 at oscillation frequency $\omega=6$ keV. 
 }
 \label{fig:AC5}
\end{figure}
%%%%%%%%%%%%%%%%%%%%%%%%%%%%%%%%%%%%%%%%%%%%%%%%%%%%%%%%%%%%%%%%%%%%%%%%%%%%%%%%

In a realistic description of nuclear fusion one has to further consider 
nuclear attraction. 
In its simplest form such an attraction can be incorporated into the system 
of equations by adding a constant depth 
$V_1 > 0$ in Eq.~\eqref{eq:truncated}.
Calculation of the transmission probability as a function of this parameter 
$V_1$ reveals a rich, non-monotonic behavior, cf.~Fig.~\ref{fig:AC5}. 

We identify three regions. 
In the limit of shallow depths $V_1<100~\rm keV$, the zeroth Floquet
channel $T_0$ corresponding to the incident energy $E$ yields the main 
contribution. 
In this regime, the total tunneling probability increases with $V_1$, 
but already before the maximum tunneling probability is reached, 
we see that this increase cannot be explained by the zeroth Floquet
channel $T_0$ alone, i.e., the side-bands ($T_{\pm1}$ and $T_{\pm2}$ etc.)
become important. 

In the intermediate regime between 100~keV and 1~MeV, the total tunneling 
probability assumes its maximum and starts to decrease with $V_1$. 
At the same time, the side-bands (first $T_{\pm1}$, then $T_{\pm2}$ etc.)
dominate the zeroth Floquet channel $T_0$.

In the limit of very deep potentials $V_1>1~\rm MeV$, 
classified as the third regime, the total tunneling probability becomes 
distributed over more and more Floquet channels (rendering the numerical
integration increasingly difficult). 
%any distinction between main- and sidebands has vanished. 
The final particle energy has effectively become continuous. 

%%%%%%%%%%%%%%%%%%%%%%%%%%%%%%%%%%%%%%%%%%%%%%%%%%%%%%%%%%%%%%%%%%%%%%%%%%%%%%%%
\subsection{Relative Enhancement} 
\label{Sec:enhance} 
%%%%%%%%%%%%%%%%%%%%%%%%%%%%%%%%%%%%%%%%%%%%%%%%%%%%%%%%%%%%%%%%%%%%%%%%%%%%%%%%

In contrast to Fig.~\ref{fig:AC5} displaying the total tunneling probability, 
let us now discuss the enhancement in comparison to the static case. 
As we have observed above, the off-set $V_1$ plays an important role.
Choosing a realistic value for $V_1$, however, is not so simple.
First, it should be remembered that $V_1$ serves as an effective toy model 
for the nuclear attraction at small distances. 
Second, our analysis is effectively one dimensional, whereas the real 
truncated Coulomb potential~\eqref{eq:truncated} is three-dimensional. 
Here, we choose $V_1=17.6~\rm MeV$ in order to accommodate for the 
energy gain in DT fusion.

\begin{figure}[t]
 \includegraphics[width = 0.49\textwidth]{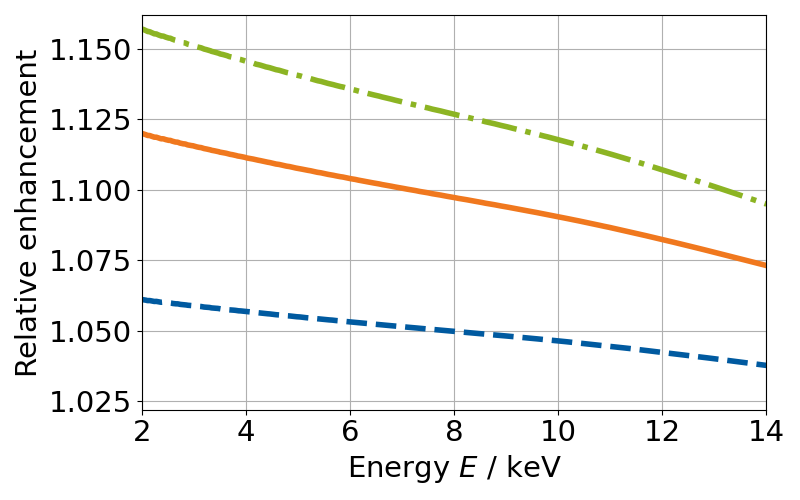}
 \caption{
 Relative enhancement in the total particle transmission probability through the truncated Coulomb barrier as a function of initial energy $E$ displayed for a variety of external field 
 strengths ${\mathfrak E}_0$ 
 (dashed blue: $5 \times 10^{16}$~V/m, % in blue, 
 solid orange: $7 \times 10^{16}$~V/m, %in orange and 
 and dashed-dotted green: $8 \times 10^{16}$~V/m). 
 Due to the presence of a strong attractive potential, $V_1=17.6$~MeV, 
 no discernible resonances around the quiver frequency $\omega = 6~\si{\kilo\electronvolt}$ 
 occur. 
 The overall enhancement effect is nevertheless about $10$~\%. 
 }
 \label{fig:AC6}
\end{figure}

The energy dependence of the relative enhancement in shown in 
Fig.~\ref{fig:AC6}, where we have chosen the interval between 2 and 14~keV 
which should be relevant for possible future technological applications. 
Consistent with the discussions in the previous sub-section, we do not 
observe any (discernible) resonances.
Nevertheless, the relative amplification is on the level of ten percent 
or more for field strengths below $10^{17}~\rm V/m$. 
For higher field strengths, it can be even stronger, see the following discussion.
For further technical details we refer to Appendix \ref{app:D}. 

%%%%%%%%%%%%%%%%%%%%%%%%%%%%%%%%%%%%%%%%%%%%%%%%%%%%%%%%%%%%%%%%%%%%%%%%%%%%%%%%
%%%%%%%%%%%%%%%%%%%%%%%%%%%%%%%%%%%%%%%%%%%%%%%%%%%%%%%%%%%%%%%%%%%%%%%%%%%%%%%%

%%%%%%%%%%%%%%%%%%%%%%%%%%%%%%%%%%%%%%%%%%%%%%%%%%%%%%%%%%%%%%%%%%%%%%%%%%%%%%%%
%%%%%%%%%%%%%%%%%%%%%%%%%%%%%%%%%%%%%%%%%%%%%%%%%%%%%%%%%%%%%%%%%%%%%%%%%%%%%%%%
\section{Conclusion} 
\label{Conclusion}
%%%%%%%%%%%%%%%%%%%%%%%%%%%%%%%%%%%%%%%%%%%%%%%%%%%%%%%%%%%%%%%%%%%%%%%%%%%%%%%%
%%%%%%%%%%%%%%%%%%%%%%%%%%%%%%%%%%%%%%%%%%%%%%%%%%%%%%%%%%%%%%%%%%%%%%%%%%%%%%%%

We have studied the enhancement of Schr\"odinger tunneling through a 
one-dimensional static potential barrier $V(x)$ induced by a time-dependent 
electric field ${\mathfrak E}(t)$ which is harmonically oscillating with 
frequency $\omega$.  
After transforming to the Kramers-Henneberger frame, we have derived and 
solved the equations for the Floquet channels which determine the transmission
and reflection amplitudes. 

As a first example, we considered the rectangular potential, which can be 
considered as a one-dimensional toy model for a scanning tunneling microscope 
(STM) or the tunneling through an insulating layer between two conductors.
Due to its simple structure, the results for the rectangular potential can be 
benchmarked with analytical approximations (e.g., in the thin-barrier as well as the 
opaque-barrier regime).  
We found resonances in the tunneling probability, e.g., 
at $E=\omega$ and $E=2\omega$, i.e., when Floquet channels open (or close).  
These resonances are pure non-adiabatic effects and clearly show that neither 
the static potential approximation nor the time-averaged potential approximation 
are applicable in these cases. 

\begin{figure}[t]
 \includegraphics[width = 0.49\textwidth]{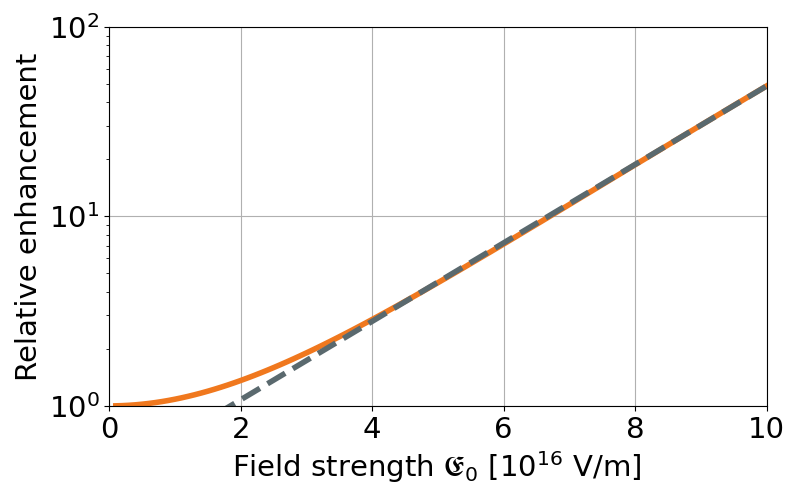}
 \caption{
 Relative enhancement in the total particle transmission probability as a 
 function of field strength ${\mathfrak E}$ for the truncated Coulomb potential
 relevant for nuclear fusion at $E = \omega = 2$ keV and $V_1=0$.
 For small field strengths, we observe an approximately quadratic growth which 
 is consistent with lowest-order perturbation theory. 
 For larger field strengths, however, the growth becomes exponential, which is due to the non-perturbative nature of tunneling and also indicating that lowest-order perturbation theory is no longer adequate.
 }
 \label{fig:Conclusion}
\end{figure}

As a second example, we investigated the truncated Coulomb potential relevant 
for nuclear fusion where we again found that the time-averaged potential 
approximation is not applicable for the parameters under consideration. 
Even though one can in principle also observe resonances in this scenario,
they are suppressed by the gradual change of the potential at the outer 
turning point and the large potential drop or off-set at the inner turning point. 
Nevertheless, for field strengths of the order of $10^{17}~\rm V/m$, 
one can observe a significant enhancement of the tunneling probability, 
see Fig.~\ref{fig:Conclusion}.

%%%%%%%%%%%%%%%%%%%%%%%%%%%%%%%%%%%%%%%%%%%%%%%%%%%%%%%%%%%%%%%%%%%%%%%%%%%%%%%%
%%%%%%%%%%%%%%%%%%%%%%%%%%%%%%%%%%%%%%%%%%%%%%%%%%%%%%%%%%%%%%%%%%%%%%%%%%%%%%%%

In conclusion, the results on field-induced quantum tunneling look promising 
and encourage further research in the field of dynamical assistance. 
Furthermore, there is an apparent universality of quantum tunneling in 
oscillating electric fields across all energy scales, so this research could 
have beneficial spin-offs in other branches of physics. 

%%%%%%%%%%%%%%%%%%%%%%%%%%%%%%%%%%%%%%%%%%%%%%%%%%%%%%%%%%%%%%%%%%%%%%%%%%%%%%%%
%%%%%%%%%%%%%%%%%%%%%%%%%%%%%%%%%%%%%%%%%%%%%%%%%%%%%%%%%%%%%%%%%%%%%%%%%%%%%%%%
\section{Outlook} 
\label{Sec:FApp}
%%%%%%%%%%%%%%%%%%%%%%%%%%%%%%%%%%%%%%%%%%%%%%%%%%%%%%%%%%%%%%%%%%%%%%%%%%%%%%%%
%%%%%%%%%%%%%%%%%%%%%%%%%%%%%%%%%%%%%%%%%%%%%%%%%%%%%%%%%%%%%%%%%%%%%%%%%%%%%%%% 

Since tunneling is ubiquitous in physics, our approach and results can be 
applied or generalized to many other cases and in several directions. 
For example, while we focused on transmission, one could also consider the 
reflection amplitudes -- such as cold neutrons scattered off a vibrating 
potential barrier. 
Not surprisingly, one also obtains resonances in this case, which have actually 
been measured in inelastic neutron scattering spectra \cite{PhysRevA.53.319}.
The fact that these experiments have been performed in the neV-regime 
further highlights the universality of our approach. 

Note that the Floquet channels in our approach are closely related to 
the mechanism of high harmonic generation, e.g., where photon beams are 
reflected off a plasma mirror 
\cite{doi.org/10.1038/nphys338, PhysRevLett.7.118}. 
In this case, the plasma mirror can be modeled as a vibrating potential 
barrier which is set in motion by the incident laser beam.
As a result, the reflected beam consists of the incident frequency $\omega$ 
plus the higher-order Floquet channels $2\omega$, $3\omega$ etc.  

As another link to existing phenomena, it would be interesting to 
study the relation or interplay between the resonance effects discussed 
here on the one hand and resonant tunneling in static (e.g., double-well)
potentials $V(x)$ on the other hand, see, e.g., 
\cite{RTD,Platero} for an overview.
The properties of these static resonances depend sensitively on the shape 
of the potential $V(x)$ and adding a field ${\mathfrak E}(t)$ 
would provide even more knobs (field strengths and driving frequency) 
for quantum control, e.g., quantum ratchets or spectral filters for  
incoming and outgoing waves separately \cite{PhysRevLett.87.070601, PhysRevLett.79.10, doi.org/10.1126/science.286.5448.231}.

Furthermore, our research may in principle also be extended to semiconductor physics. Electrons in semiconductors are characterized by an effective mass, which can be significantly smaller than the free electron mass of $511$ keV. Consequently, the energy and field strengths required to observe field-assisted quantum tunneling are substantially reduced. For instance, for a rectangular barrier and an effective mass of $50$ keV the respective field frequency for dynamically assisted transmission is in the range of $10$ meV and the field strength is $\sim 5 \times 10^6$ V/m.
 
In atomic physics, a prominent example of tunneling is the ionization
by a strong and slowly varying electric field (field ionization), see, 
e.g., Refs.~\cite{doi.org/10.1088/0022-3700/18/8/001,PhysRevLett.63.2212,Hartung,PhysRevA.91.031402,KeldyshIO}.
The enhancement (or suppression) of this process by an additional 
time-dependence (e.g., a time-dependent electric field) can also be 
studied by a suitable generalization of the approach presented here. 
Going to much shorter length and higher energy scales, quite analogous 
phenomena occur in nuclear physics, see, e.g.,
\cite{Kelkar, PhysRevLett.108.142501, PhysRevC.102.064629, PhysRevC.106.064610}. 
As one example, the dynamical assistance of $\alpha$-decay has been 
discussed (controversially) in Ref. \cite{PP20,BaiDeng, PhysRevLett.119.202501, Szilvasi, Rizea1, Rizea2}. 

In general, the study of assisted quantum tunneling in three dimensions is the next objective. Then, the angular momentum of the incoming wave function and, consequently, the angular momentum barrier has to be taken into account. One might turn this into an advantage as different transitions have to be considered, such as from an incoming $l=1$-state to an $l=0$-state, where only the latter can effectively tunnel through the composite nuclear barrier.

Apart from the case of non-relativistic tunneling investigated here,
one can also consider relativistic tunneling. 
Prominent examples include false vacuum decay in early cosmology \cite{PhysRevD.15.2929, PhysRevD.16.1762,PhysRevLett.123.031601,PhysRevA.105.L041301}
and the Sauter-Schwinger effect, i.e., electron-positron pair creation 
out of the vacuum by a strong electric field \cite{sauter_1931, schwinger_1951, PhysRevLett.101.130404,PhysRevLett.129.241801}. 
Even though the underlying (Dirac or Klein-Fock-Gordon) equations are 
different from the Schr\"odinger equation studied here, one faces the 
same problem that tunneling in space-time dependent backgrounds is not 
fully understood yet, see also Refs.~\cite{Ilderton:2021zej, Diez:2022ywi, Alvarez-Dominguez:2023zsk, PhysRevA.73.022114, PhysRevLett.92.040406, PhysRevA.85.033625}.
On a completely different scale, these phenomena could also be related 
to the pseudo-relativistic dynamics in graphene \cite{doi.org/10.1126/science.abi8627, Schmitt:2022pkd, Biswas, PhysRevB.78.165420,arXiv:2305.04895}.

%%%%%%%%%%%%%%%%%%%%%%%%%%%%%%%%%%%%%%%%%%%%%%%%%%%%%%%%%%%%%%%%%%%%%%%%%%%%%%%%
%%%%%%%%%%%%%%%%%%%%%%%%%%%%%%%%%%%%%%%%%%%%%%%%%%%%%%%%%%%%%%%%%%%%%%%%%%%%%%%% 
\acknowledgments
%%%%%%%%%%%%%%%%%%%%%%%%%%%%%%%%%%%%%%%%%%%%%%%%%%%%%%%%%%%%%%%%%%%%%%%%%%%%%%%%
%%%%%%%%%%%%%%%%%%%%%%%%%%%%%%%%%%%%%%%%%%%%%%%%%%%%%%%%%%%%%%%%%%%%%%%%%%%%%%%%

Calculations have been performed on the Hemera cluster in Dresden-Rossendorf. 
D.R.~wants to thank Falk Adamietz for inspiring discussions. 
Funded by the Deutsche Forschungsgemeinschaft (DFG, German Research Foundation) 
-- Project-ID 278162697 -- SFB 1242. 
This research was supported in part by Perimeter Institute for Theoretical 
Physics. 
Research at Perimeter Institute is supported by the Government of Canada through 
the Department of Innovation, Science and Economic Development and by the 
Province of Ontario through the Ministry of Research and Innovation.
%
%\key{Weitere Textbausteine?}

%%%%%%%%%%%%%%%%%%%%%%%%%%%%%%%%%%%%%%%%%%%%%%%%%%%%%%%%%%%%%%%%%%%%%%%%%%%%%%%%
%%%%%%%%%%%%%%%%%%%%%%%%%%%%%%%%%%%%%%%%%%%%%%%%%%%%%%%%%%%%%%%%%%%%%%%%%%%%%%%%
\appendix
%%%%%%%%%%%%%%%%%%%%%%%%%%%%%%%%%%%%%%%%%%%%%%%%%%%%%%%%%%%%%%%%%%%%%%%%%%%%%%%%
%%%%%%%%%%%%%%%%%%%%%%%%%%%%%%%%%%%%%%%%%%%%%%%%%%%%%%%%%%%%%%%%%%%%%%%%%%%%%%%%

%%%%%%%%%%%%%%%%%%%%%%%%%%%%%%%%%%%%%%%%%%%%%%%%%%%%%%%%%%%%%%%%%%%%%%%%%%%%%%%%
%%%%%%%%%%%%%%%%%%%%%%%%%%%%%%%%%%%%%%%%%%%%%%%%%%%%%%%%%%%%%%%%%%%%%%%%%%%%%%%%
\section{Theoretical Formalism} 
\label{app:A}
%%%%%%%%%%%%%%%%%%%%%%%%%%%%%%%%%%%%%%%%%%%%%%%%%%%%%%%%%%%%%%%%%%%%%%%%%%%%%%%%
%%%%%%%%%%%%%%%%%%%%%%%%%%%%%%%%%%%%%%%%%%%%%%%%%%%%%%%%%%%%%%%%%%%%%%%%%%%%%%%%

In this section we want to detail the building blocks of the theoretical framework. This includes a full discussion of the channel equation formalism itself, as well as further explanations of how we incorporated the Kramers-Henneberger frame or a resonance study into it. Nevertheless, most of the basic derivations can be found in Refs.~\cite{doi.org/10.1002/qua.560160606, doi.org/10.1016/0039-6028(82)90711-7} and, in more detail, in Ref.~\cite{Razavy}.

\subsection{The Kramers-Henneberger frame} 
\label{Sec:KH}

The transformation from laboratory frame to Kramers-Henneberger frame is performed by means of the Kramers-Henneberger transformation. It provides an exact mapping between a vector potential $A(t)$ and a wave function's quiver motion $\chi(t)$. Starting point is the Schrödinger equation
\begin{equation}
 \I \hbar\frac{\partial}{\partial t} \psi(x,t) = \left( \frac{\Big( p - q A(t) \Big)^2}{2\mu} + V(x) \right) \psi(x,t),
 \label{eq:Schr}
\end{equation}
with a purely time-dependent vector potential $A(t)$ and time-independent scalar potential $V(x)$. The quadratic term $\dot \xi(t)= q^2 A(t)^2/(2\mu)$ can be eliminated by means of a transformation 
\begin{equation}
 \psi(x,t) \to \exp \Big( \I \xi(t) \Big) \psi(x,t), 
\end{equation}
while the linear term $\chi(t)$ can be incorporated in a spatial quiver motion $\dot \chi(t) = q/\mu A(t)$
through application of the translation operator
\begin{equation}
 \psi(x,t) \to \exp \Big( \I \chi(t) p \Big) \psi(x,t),   
 \label{eq:trans}
\end{equation}
and switching to a moving frame of reference
\begin{equation}
 x \to x + \chi(t), \quad \text{thus} \quad \partial_t \to \partial_t - \dot \chi(t) \partial_x.
\end{equation}
In this way we obtain the Schrödinger equation in the Kramers-Henneberger frame
\begin{equation}
 \I \hbar\frac{\partial}{\partial t} \psi_{\rm KH}(x,t) = \Big( \frac{p^2}{2\mu} + V \big(x - \chi(t) \big) \Big) \psi_{\rm KH}(x,t).
\end{equation}

\subsection{General formalism for uneven asymptotic levels}
\label{Sec:Framework}

We want to go into detail regarding the derivation of the system of coupled equations that has been rather superficially presented in the main text. This is done in order to have a coherent, self-contained paper.

In order to employ such potential of arbitrary asymptotic levels we first have to tweak the basic formalism slightly. More specifically, we write for the potential in the Kramers-Henneberger frame
\begin{align}
 V(x,t) = W(x,t) - V_1 \Theta(x), 
\end{align}
where $V(x\rightarrow \infty,t)=-V_1$ and $V(x\rightarrow -\infty,t)=0$, such that $W(x\rightarrow\pm\infty,t)=0$.  We thus split the potential into a dynamical potential $W(x,t)$ and a step function $V_1 \Theta(x)$. For the sake of clarification, in the main text we only showed the simple derivation for a potential where $V_1=0$.

The Schrödinger equation in the Kramers-Henneberger frame is therefore given by
\begin{multline}
 \I \hbar \frac{\partial}{\partial t} \psi_{\rm KH}(x,t) = \\ \bigg[-\frac{1}{2 \mu} \frac{\partial^2}{\partial_x^2} + W \big(x - \chi(t) \big) - V_1\Theta(x) \bigg] \psi_{\rm KH}(x,t).
\end{multline}
Within Floquet theory the wave function as well as the potentials are decomposed into Fourier modes
\begin{align}
 & \psi_{\rm KH}(x,t) && = \ee^{-\I Et} \times \sum_{n=-\infty}^{\infty} u_n(x) \ee^{\I n\omega t},\\
 & W(x,t) && = \phantom{\ee^{-\I Et} \times} \sum_{n=-\infty}^{\infty} W_n (x)\,\ee^{\I n\omega t},
\end{align}
so that we arrive at the channel equation
\begin{equation}
 \frac{{\rm d}^2}{{\rm d} x^2} u_n(x) + k_{n}^{2}(x) u_n(x) = \sum_{m=-\infty}^{\infty} w_{nm}(x) u_m(x),
 \label{eq:form}
\end{equation}
with $k_n(x) \equiv k_n=\sqrt{2 \mu (E + n\omega)}$ for $x<0$ and $k_n(x) \equiv k^{V_1}_n=\sqrt{2 \mu (E -V_1+ n\omega)}$ for $x\geq 0$. Furthermore, we have $w_{nm}(x) \equiv 2 \mu \,W_{n-m}(x)$.

The formal solution to Eq.~\eqref{eq:form} is given by 
\begin{multline}
 u_{nl}(x) = \phi^n_2(x) \delta_{nl} + \\
 \int_{-\infty}^\infty {\rm d}x' G_n(x,x') \times \sum_m w_{nm}(x') u_{ml}(x'),
 \label{eq:gen}
\end{multline}
with Green's function
\begin{align}
G_n(x,x') = C_n 
    \begin{cases}
        \phi^n_1(x) \phi^n_2(x') & \text{for } x<x' , \\
        \phi^n_2(x)\phi^n_1(x') & \text{for } x>x' ,
    \end{cases}
\end{align}
and $C_n= -\I / \left( k_n+k_n^{V_1} \right)$.

The free wave functions, in turn, are given by
\begin{align}
&\phi^n_1(x) = \ee^{-\I k_n x} \Theta(-x) + \\
& \left\{ \frac{1}{2} \left( 1-\frac{k_n}{k_n^{V_1}} \right) \ee^{\I k_n^{V_1} x}
+       \frac{1}{2} \left( 1+\frac{k_n}{k_n^{V_1}} \right) \ee^{-\I k_n^ x} \right\} \Theta(x), \notag \\
&\phi^n_2(x) = \ee^{\I k^{V_1}_n x} \Theta(x) + \\
& \left\{ \frac{1}{2} \left( 1+\frac{k^{V_1}_n}{k_n} \right) \ee^{\I k_n x}
+       \frac{1}{2} \left( 1-\frac{k_n^{V_1}}{k_n} \right) \ee^{-\I k_n x} \right\} \Theta(-x). \notag
\end{align}

In the asymptotic limits we find the solutions
\begin{align}
u_{nl}(x\rightarrow-\infty) &= \phi^n_2(x) \delta_{nl} + \rho_{nl} \phi^n_1(x), \\
u_{nl}(x\rightarrow+\infty) &= \phi^n_2(x) \delta_{nl} + \tau_{nl} \phi^n_2(x),
\end{align}
where
\begin{align}
 \rho_{nl} &=C_n
 \int_{-\infty}^\infty {\rm d}x' \phi^n_2(x') \times \sum_m w_{nm}(x')
u_{ml}(x'), \\
 \tau_{nl} &= C_n
 \int_{-\infty}^\infty {\rm d}x' \phi^n_1(x') \times \sum_m w_{nm}(x')
u_{ml}(x').
\end{align}
Rather than solving for the wavefunction $u_{nl}(x)$, we will 
derive a system of differential equations from which $\tau_{nl}$ and $\rho_{nl}$ can be obtained directly.
To do this, we first generalise the formal solution \eqref{eq:gen} to
\begin{align}
u_{nl}(x,y)&=\phi_2^{n}(x)\delta_{nl} + \label{eq:formalsol} \\
&\int^{\infty}_y {\rm d}x' \ G_{n}(x,x')
\sum_{m}w_{n m}(x')u_{ml}(x',y). \notag
\end{align}
By taking the derivative with respect to $y$ and multiplying Eq.~\eqref{eq:formalsol} with a yet undetermined matrix $B$ we then obtain the relation
\begin{align}
\label{eq:formalsol2}
&\sum_l \frac{\partial u_{nl}(x,y)}{\partial y}B_{ls}(y)=\\
&-\sum_{l,m}C_n \phi_2^n(x)\phi_1^n(y)w_{nm}(y) u_{ml}(y,y)B_{ls}(y)\nonumber\\
&
+ \int^{\infty}_y {\rm d}x' \ G_{n}(x,x')
\sum_{l,m}w_{n m}(x')\frac{\partial u_{ml}(x',y)}{\partial y}B_{ls}(y)\nonumber.
\end{align}
If we require the newly introduced matrix $B$ to satisfy
\begin{align}
-\sum_{l,m} C_n \phi_1^n(y)w_{nm}(y) u_{ml}(y,y)B_{ls}(y)=\delta_{ns},
\label{eq:matrixB}
\end{align}
we find that Eq.~\eqref{eq:formalsol2} is equivalent to Eq.~\eqref{eq:formalsol} under the relation
\begin{align}
\sum_l \frac{\partial u_{ml}(x,y)}{\partial y}B_{ls}(y)=u_{ms}(x,y).
\end{align}
In the next step, we consider the generalized amplitudes
\begin{align}
\rho_{nl}(y)&= C_n\int^{\infty}_y {\rm d}x' \ \phi_2^n(x') \sum_m w_{nm}(x') u_{ml}(x',y),\\
\tau_{nl}(y)&= C_n\int^{\infty}_y {\rm d}x' \ \phi_1^n(x') \sum_m w_{nm}(x') u_{ml}(x',y) .
\end{align}
Taking the derivative with respect to $y$ and using the relation stated in Eq.~\eqref{eq:matrixB}, we 
obtain
\begin{align}
& \frac{{\rm d} \rho_{nl}(y)}{{\rm d}y} = \label{odeR} \\
& -C_n \phi_2^n(y) \sum_m w_{nm}(y) u_{ml}(y,y)+\sum_s \rho_{ns}(y)B^{-1}_{sl}(y), \notag \\
& \frac{{\rm d}\tau_{nl}(y)}{{\rm d}y} = \label{odeT} \\
& -C_n \phi_1^n(y) \sum_m w_{nm}(y) u_{ml}(y,y)+\sum_s \tau_{ns}(y)B^{-1}_{sl}(y). \notag 
\end{align}
The inverse of the matrix $B$ reads
%The inverse of the matrix $B$ is determined from Eq.~\eqref{eq:matrixB}
%
\begin{align}
 %B^{-1}_{lm}(y)=-C_l\, \phi_1^l(y)\sum_s w_{ls}(y) u_{sm}(y,y),
 B^{-1}_{sl}(y)=-C_s\, \phi_1^s(y)\sum_m w_{sm}(y) u_{ml}(y,y),
 \label{inverseB}
\end{align}
whereas from the formal solution \eqref{eq:formalsol} we obtain
\begin{align}
\label{eq:formalsol3}
u_{ml}(y,y) = \phi_2^{m}(y) \delta_{ml} + \phi_1^m(y) \rho_{ml}(y) .
%u_{nl}(y,y)=\phi_2^{n}(y)\delta_{nl}+\phi_1^n(y)\rho_{nl}(y).
\end{align}
Inserting these identities \eqref{inverseB}-\eqref{eq:formalsol3} into the set of differential 
equations \eqref{odeR}-\eqref{odeT} we finally find 
\begin{widetext}
\begin{align}
\frac{{\rm d} \rho_{nl}(y)}{{\rm d}y}=-\sum_{s,m}C_s
\left[\phi_2^s(y)\delta_{ns} +\rho_{ns}(y)\phi_1^s(y)\right]w_{sm}(y)\left[\phi_2^m(y)\delta_{ml}+\rho_{ml}(y)\phi_1^m(y)\right], \\
%\frac{{\rm d} \rho_{nl}(y)}{{\rm d}y}=-\sum_{s,p}C_s
%\left[\delta_{sn}\phi_2^s(y)+\rho_{ns}(y)\phi_1^s(y)\right]w_{sp}(y)\left[\delta_{pl}\phi_2^p(y)+\rho_{pl}(y)\phi_1^p(y)\right], \\
%
\frac{{\rm d} \tau_{nl}(y)}{{\rm d}y}=-\sum_{s,m}C_s
\left[\phi_1^s(y)\delta_{ns}+\tau_{ns}(y)\phi_1^s(y)\right]w_{sm}(y)\left[\phi_2^m(y)\delta_{ml}+\rho_{ml}(y)\phi_1^m(y)\right]. 
%\frac{{\rm d} \tau_{nl}(y)}{{\rm d}y}=-\sum_{s,p}C_s
%\left[\delta_{sn}\phi_1^s(y)+\tau_{ns}(y)\phi_1^s(y)\right]w_{sp}(y)\left[\delta_{pl}\phi_2^p(y)+\rho_{pl}(y)\phi_1^p(y)\right]. 
\end{align}
\end{widetext}
This set of equations has to be integrated from $y \to \infty$ to $y \to - \infty$. Initial conditions are given by 
\begin{equation}
 \rho_{nl}(y\rightarrow\infty) = 0, \quad \tau_{nl}(y\rightarrow\infty) = 0.
\end{equation}
Reflection and transmission amplitudes are identified as
\begin{align}
R_{nl} &= \rho_{nl}(y\rightarrow-\infty), \\
T_{nl}=\delta_{nl} + \tau_{nl} &= \delta_{nl} + \tau_{nl}(y\rightarrow-\infty).
\end{align}

In the presentation in the main text we opted for a more comprehensive representation that is easier to follow. In order to recover the simplified version, we have to have a potential that truly vanishes asymptotically as then $k_s^{V_1}$ is equal $k_s$, thus $C_s=-\I/\left( k_s+k_s^{V_1} \right)$, and $w_{sm}(y)$ is reduced to $v_{sm}(y)$. For the wave functions $\phi_1$ and $\phi_2$ an expansion in terms of plane waves further yields
$\phi_2^s(y) = \ee^{\I k_s y}$ and $\phi_1^s(y) = \ee^{-\I k_s y}$, respectively, completing the translation.

\subsection{Resonances}
\label{Sec:Resonances}

In order to confirm the presence of resonances in the transmission probability we rely on techniques developed in a non-hermitian formulation of quantum mechanics, see Ref.~\cite{Moiseyev} for an overview.
For our study, we search for solutions in the Floquet-Schrödinger equation \eqref{eq:u} by performing a scaling transformation $x \to x \, \ee^{\I \theta}$, where $\theta$ is a real constant, 
\begin{multline}
\frac{\rm d^2}{{\rm d} x^2} u_n(x) + 2 \mu ({\cal E}+n\omega) u_n(x) = \\
\sum_{m=-\infty}^{\infty} v_{nm}(x) u_m(x).
\end{multline}
In this case, $v_{nm}(x) \equiv 2 \mu \,V_{n-m}(x)$, which expands the solution space to the complex plane.
Consequently, the only normalizable solutions in energy are to be found in the complex plane. We indicate this shift by writing ${\cal E}$ instead of $E$. 

As a matter of fact we expect the complex solutions ${\cal E}$ to be localized either along straight lines (under an angle $\alpha$ depending on the chosen value of $\theta$ measuring the tilt from the real x-axis) or at isolated points. To better visualize the former aspect conceptually, we want to consider the limit of a decoupled set of equations, that is where $v_{nm}=0$. In this case, the Floquet-Schrödinger equation reduces to a simple eigenwert equation
\begin{equation}
 \left( - \frac{\ee^{-2\I \theta}}{2 \mu} \frac{{\rm d}^2}{{\rm d}x^2} - n\omega \right) u_n(x)  = {\cal E} u_n(x).   
\end{equation}
Solutions in $u_n(x)$ are of the form $\sim \exp \left(\I k x \ee^{-\I \theta} \right)$ and demanding these eigenvectors to be normalizable we find the necessary condition $k=\lvert k \rvert \ee^{\I \theta}$. Consequently, the eigenvalues are given by
\begin{equation}
 {\cal E} = -n \omega + \lvert k \rvert^2 \ee^{-2\I \theta}.
 \label{eq:Eigen}
\end{equation}
Resonances only appear if the wave function is coupled to a quivering potential, hence not all $v_{nm}$ vanish. Solutions of ${\cal E}$ that do not follow the simple relation \eqref{eq:Eigen} can therefore be associated with the resonant enhancement we see in the transmission probability.

\begin{figure}[t]
 \includegraphics[width=0.49\textwidth]{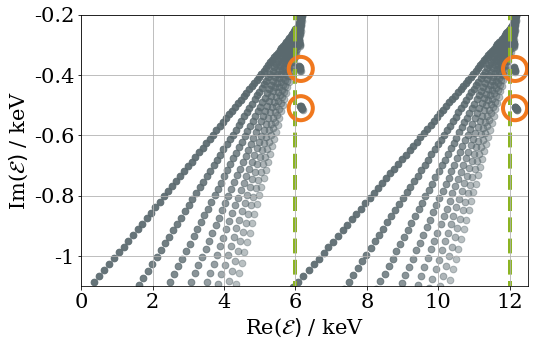}
 \caption{
 Eigenvalue structure of the Schrödinger equation for the Coulomb potential in the complex energy-plane ${\cal E}$ for different values of $\theta$. The singular points, highlighted through orange circles, mark the onset of a resonance. These locations coincide with the frequency of the applied field $\omega$ (green, dashed lines) which, for this specific example, was chosen to be $\omega=6$ keV. The grey lines mark branch cuts which simply refer to the regular spectrum in the continuum. The field strength was fixed at ${\mathfrak E}_0 \approx 3\times 10^{16}$ V/m.
 }
 \label{fig:AppAC1}
\end{figure}

In Fig.~\ref{fig:AppAC1} an exemplary plot of the imaginary eigenvalue landscape of the shifted Schrödinger equation is displayed. Of interest are the locations of the isolated points as they mark the existence of resonances. Such isolated points appear at real energies of zero, $6$ keV, $12$ keV, and so on. They are also (almost) independent of the shift parameter $\theta$ in contrast to the regular spectrum which builds out tilted line structures.

%%%%%%%%%%%%%%%%%%%%%%%%%%%%%%%%%%%%%%%%%%%%%%%%%%%%%%%%%%%%%%%%%%%%%%%%%%%%%%%%
%%%%%%%%%%%%%%%%%%%%%%%%%%%%%%%%%%%%%%%%%%%%%%%%%%%%%%%%%%%%%%%%%%%%%%%%%%%%%%%%

%%%%%%%%%%%%%%%%%%%%%%%%%%%%%%%%%%%%%%%%%%%%%%%%%%%%%%%%%%%%%%%%%%%%%%%%%%%%%%%%
%%%%%%%%%%%%%%%%%%%%%%%%%%%%%%%%%%%%%%%%%%%%%%%%%%%%%%%%%%%%%%%%%%%%%%%%%%%%%%%%
\section{Rectangular Barrier} 
\label{app:C}
%%%%%%%%%%%%%%%%%%%%%%%%%%%%%%%%%%%%%%%%%%%%%%%%%%%%%%%%%%%%%%%%%%%%%%%%%%%%%%%%
%%%%%%%%%%%%%%%%%%%%%%%%%%%%%%%%%%%%%%%%%%%%%%%%%%%%%%%%%%%%%%%%%%%%%%%%%%%%%%%%

In this section the solutions for field-induced quantum tunneling through a rectangular barrier are presented. We also give an outlook regarding an analytical solution for such tunneling with an electric field present.

\subsection{Static Barrier}
\label{Sec:Rec0}

Obtaining the transmission rates for particle tunneling through a rectangular barrier is one of the textbook examples of quantum mechanics. If, however, the potential in front and behind the barrier is not on the same level the results for the transmission rates are already much less well known. For the sake of providing a complete picture of our analysis on field-assisted quantum tunneling we thus state the overall strategy for finding analytical solutions and also provide various example results, e.g., for a rectangular barrier on uneven ground.

In terms of scattering at a rectangular barrier we find three regions 
\begin{equation}
V(x) = 
 \begin{cases} 
  0 & \text{for } x \leq 0, \\ 
  V_0>0 & \text{for } 0 \leq x \leq L, \\ 
  -V_1<0 & \text{for } x \geq L,
 \end{cases}
\label{eq:RecV} 
\end{equation}
which we will denote according to their appearance in Eq. \eqref{eq:RecV} as region I, region II and region III, respectively. Equivalently to the textbook case, we make an ansatz for the wave function dividing it into three segments with a left- and right-moving part in each segment. The main difference is that we have three distinct wave numbers $k_{\rm I}$, $k_{\rm II}$ and $k_{\rm III}$, one for each region.
For our purposes we demand $-V_1 < 0 < E < V_0$ and have the wave function propagate from left to right. Then, we obtain for reflection $r_E$ and transmission $t_E$ rates
\begin{widetext}
\begin{align}
 r_E &= 
\frac{(k_{\rm I}-k_{\rm II}) (k_{\rm II}+k_{\rm III})+(k_{\rm I}+k_{\rm II}) (k_{\rm II}-k_{\rm III}) e^{2 \I k_{\rm II} L}}{(k_{\rm I}+k_{\rm II}) (k_{\rm II}+k_{\rm III})+(k_{\rm I}-k_{\rm II}) (k_{\rm II}-k_{\rm III}) \ee^{2 \I k_{\rm II} L}}, \\
 t_E &= 
 -\frac{4 k_{\rm I} k_{\rm II} \ee^{\I L (k_{\rm II}-k_{\rm III})}}{-(k_{\rm I}+k_{\rm II}) (k_{\rm II}+k_{\rm III})+(k_{\rm II}-k_{\rm I}) (k_{\rm II}-k_{\rm III}) \ee^{2 \I k_{\rm II} L}},    
\end{align}
\end{widetext}
with 
\begin{multline}
 k_{\rm I} = \sqrt{2 \mu E}, \, k_{\rm II} = \sqrt{2 \mu (E-V_0)}\,\\ \text{and} \, k_{\rm III} = \sqrt{2 \mu (E+V_1)}.
\end{multline}
Reflection $R$ and transmission probabilities $T$ are then obtained through conservation of the total probability current
\begin{equation}
 1 = R + T = \lvert r_E \rvert^2 + \sqrt{1+V_1/E} \, \lvert t_E \rvert^2  .  
\end{equation}
Note that the prefactor $\sqrt{1+V_1/E}$ increases with a larger potential depth in region III. If $ V_1 \gg E$ the wave is thus effectively reflected at the second boundary. 

\subsection{Quivering barrier}
\label{Sec:Rec1}

In this section we consider tunneling through a quivering rectangular potential, cf. Eq. \eqref{eq:RecV}.
Due to its special shape, the potential is locally constant in each region, an analytical solution for the reflection and transmission rate can even be found for tunneling with an additional oscillating electric field present. The procedure, while more involved, is the same as for the textbook example of tunneling through a static barrier, see also the derivation above. Nevertheless, we present the underlying concept in the following. 

The starting point is again the Schrödinger equation, cf. Eq. \eqref{eq:Schr},
\begin{equation}
 \I \hbar\frac{\partial}{\partial t} \psi(x,t) = \left( \frac{\Big( p - q A(t) \Big)^2}{2 \mu} + V(x) \right) \psi(x,t),
\end{equation}
with vector potential $A(t)$.

As $A(t)=-{\mathfrak E}_0/\omega \sin(\omega t)$ is periodic, so is the quiver motion $\chi(t) = q{\mathfrak E}_0/\mu \omega^2 \cos (\omega t)$ with amplitude $\chi_0 = q{\mathfrak E}_0/\mu \omega^2$ and, in turn, the translation operator.
Then, when applying the translation operator to make the transition to a moving frame the solutions to the Schrödinger equations are again provided in terms of region-specific wave functions,
\begin{align}
&\psi_{\rm I} &&=&& \phantom{\sum_m A^r_m} \ee^{-\I Et + i k_{{\rm I} 0} X(t)} \label{eq:App_psi1} \\
& && && + \sum_m A^l_m \ee^{-\I (E+m \omega) t - \I k_{{\rm I} m} X(t)}, \notag \\
&\psi_{\rm II} &&=&& \sum_m B^r_m \ee^{-\I (E+m \omega) t + \I k_{{\rm II} m} X(t)} \\
& && && +\sum_m B^l_m \ee^{-\I (E+m \omega) t - \I k_{{\rm II} m} X(t)}, \notag \\
&\psi_{\rm III} &&=&& \sum_m C^r_m \ee^{-\I (E+m \omega) t + \I k_{{\rm III} m} X(t)},
\label{eq:App_psi3}
\end{align}
with $X(t) = x+\chi(t)$. Each wave function is given in terms of their Fourier decomposition and with respect to the general time evolution in the Schrödinger equation 
\begin{equation}
 \psi(x,t) = \ee^{-\I Et} \sum_{n=-\infty}^{\infty} \phi_m(x) \ee^{\I m\omega t},
 \label{eq:App_phi}
\end{equation}
with generic $\phi(x)$ used as a stand-in for the left- and right-propagating waves in all three regions. If the amplitude of the quivering motion is set to zero all $\phi_m(x)$ with $m \neq 0$ vanish. Thus, the time translation factors out and the familiar structure for static tunneling is recovered as wave vectors simplify
\begin{multline}
 k_{{\rm I} m} = \sqrt{2 \mu (E+m\omega)}, \, k_{{\rm II} m} = \sqrt{2 \mu(E+m\omega-V_0)}\,\\ 
 \text{and} \quad k_{{\rm III} m} = \sqrt{2 \mu(E+m\omega+V_1)}.
\end{multline}
Also note that in the above ansatz we have already incorporated appropriate initial conditions, in particular, there is one and only one incoming mode $A^r_0=1$, thus $A^r_{m\neq0}=0$ and $C^l_m=0$. 

Furthermore, due to the fact that the potential oscillates we can use the identity 
\begin{equation}
 \ee^{a \cos(\omega t)} = \sum_n f_n(a) \ee^{\I n \omega t},    
\end{equation}
for the $\chi$-dependent part of the wave functions. This allows us to use the relation
\begin{equation}
 f_n(a)= \frac{\omega}{2\pi}\int_0^{2\pi/\omega} {\rm d}t \, \ee^{a \cos(\omega t)- \I n \omega t} = I_n(a),
\end{equation}
where $I_n(a)$ are Bessel functions of the first kind.

Junction conditions at $x=0$ and $x=L$ then yield the algebraic equations
\begin{multline}
 I_n(\I k_{{\rm I} 0} \chi_0) + \sum_m I_{n+m} (-\I k_{{\rm I} m} \chi_0) A^l_m = \\
 \sum_m I_{n+m} (\I k_{{\rm II} m} \chi_0) B^r_m + \\
 \sum_m I_{n+m} (-\I k_{{\rm II} m} \chi_0) B^l_m, 
\end{multline}
\begin{multline}
 \I k_{{\rm I} 0} I_n(\I k_{{\rm I} 0} \chi_0) - \sum_m \I k_{{\rm I} m} I_{n+m} (-\I k_{{\rm I} m} \chi_0) A^l_m = \\
 \sum_m \I k_{{\rm II} m} I_{n+m} (\I k_{{\rm II} m} \chi_0) B^r_m - \\
 \sum_m \I k_{{\rm II} m} I_{n+m} (-\I k_{{\rm II} m} \chi_0) B^l_m,
\end{multline}
\begin{multline}
 \sum_m I_{n+m} (\I k_{{\rm II} m} \chi_0) B^r_m \ee^{\I k_{{\rm II} m} L} + \\
 \sum_m I_{n+m} (-\I k_{{\rm II} m} \chi_0) B^l_m \ee^{-\I k_{{\rm II} m} L} = \\
 \sum_m I_{n+m} (\I k_{{\rm III} m} \chi_0) C^r_m \ee^{\I k_{{\rm III} m} L},
\end{multline}
\begin{multline}
 \sum_m \I k_{{\rm II} m} I_{n+m} (\I k_{{\rm II} m} \chi_0) B^r_m \ee^{\I k_{{\rm II} m} L} - \\
 \sum_m \I k_{{\rm II} m} I_{n+m} (-\I k_{{\rm II} m} \chi_0) B^l_m \ee^{-\I k_{{\rm II} m} L} = \\
 \sum_m \I k_{{\rm III} m} I_{n+m} (\I k_{{\rm III} m} \chi_0) C^r_m \ee^{\I k_{{\rm III} m} L}.
\end{multline}
This system of equations has then to be solved for the amplitudes $A^l$, $B^l$, $B^r$ and $C^l$. It is often more illuminating, however, to calculate the reflection and tunneling probabilities instead of the respective coefficients. Requiring conservation of the probability current
\begin{equation}
 j = \frac{\I}{2} \frac{\omega}{2\pi}\int_0^{2\pi/\omega} {\rm d}t
 \left( \psi \frac{\partial}{\partial_x} \psi^\ast - \psi^\ast \frac{\partial}{\partial_x} \psi \right) 
\end{equation}
yields conservation of probability (normalized to one)
\begin{equation}
 1 = \sum_{m} \frac{k_{{\rm I} m}}{k_{{\rm I} 0}} \left\lvert A^l_m \right\rvert^2 + \sum_{m} \frac{k_{{\rm III} m}}{k_{{\rm I} 0}} \left\lvert C^r_m \right\rvert^2,  
% 1 = \sum_{\substack{m \\ E + m\omega \geq 0}} \frac{k_{{\rm I} m}}{k_{{\rm I} 0}} \lvert A^l_m \rvert^2 + \sum_{\substack{m \\ E + m\omega - V_1 \geq 0}} \frac{k_{{\rm III} m}}{k_{{\rm I} 0}} \lvert C^r_m \rvert^2.  
\end{equation}
where the sums are over all real $k_{{\rm I} m}$, $k_{{\rm III} m}$ only.

\subsection{Perturbative Analysis} 
\label{Sec:Rec2}

In this section we discuss field-induced quantum tunneling through a rectangular barrier in terms of perturbation theory. This is a follow-up of the derivation in the section above. The small (perturbative) parameter is $k \chi_0$, where $k$ are again the momentum vectors in the different regions, respectively,
and $\chi_0$ is the quiver amplitude. Note that for this specific analysis we have equal asymptotic potential levels. Therefore, we assume small displacements and a barrier that is moderately high compared to the incident energy; $V_0$ in relation to $E$. In these conditions the arguments in the various Bessel functions $I_n$ are small, in particular, $I_0(\I k \chi_0) \approx 1$ and $I_1(\I k \chi_0) \approx \I k \chi_0/2$. Moreover, the mode with energy $E$ is dominant in all regions, $\lvert \phi_0 \rvert^2 \gg \lvert \phi_{\pm 1} \rvert^2$ \eqref{eq:App_phi}.

Performing a perturbative expansion of the wave functions \eqref{eq:App_psi1}-\eqref{eq:App_psi3} to zeroth order yields simplified junction conditions at $x=0$,
\begin{align}
& 1+A_0^l &= B_0^r + B_0^l \label{eq:App_zero1}, \\
& \I k_{{\rm I} 0} \left( 1- A_0^l \right) &= \I k_{{\rm II} 0} \left( B_0^r + B_0^l \right), 
\end{align}
and $x=L$,
\begin{equation}
 B_0^r \ee^{\I k_{{\rm II} 0} L} + B_0^l \ee^{- \I k_{{\rm II} 0} L} = \ee^{\I k_{{\rm I} 0} L} C_0^r, 
\end{equation}
\begin{multline}
 \I k_{{\rm II} 0} \left( B_0^r \ee^{\I k_{{\rm II} 0} L} - B_0^l \ee^{-\I k_{{\rm II} 0} L} \right) = \\
 \qquad \I k_{{\rm I} 0} \ee^{\I k_{{\rm I} 0} L} C_0^r. \label{eq:App_zero4}      
\end{multline}
In the next-to-leading-order expansion we find
\begin{equation}
 A_{\pm 1}^l = B_{\pm 1}^r + B_{\pm 1}^l, 
\end{equation}
\begin{multline}
 -\I k_{{\rm II} \pm 1} A_{\pm 1}^l = \\
 \mu V_0 \chi_0 \left( B_0^r + B_0^l \right) + k_{{\rm II} \pm 1} \left( B_{\pm 1}^r - B_{\pm 1}^l \right), 
\end{multline}
\begin{multline}
 C_{\pm 1}^r \ee^{\I k_{{\rm I} \pm 1} L} = \\
 B_{\pm 1}^r \ee^{\I k_{{\rm II}  \pm 1} L} + B_{\pm 1}^l \ee^{-\I k_{{\rm II}  \pm 1} L}, 
\end{multline}
\begin{multline}
 \I k_{{\rm I} \pm 1} C_{\pm 1}^r \ee^{\I k_{{\rm I} \pm 1} L} = \\ 
 \mu V_0 \chi_0 \left( B_0^r \ee^{\I k_{{\rm II} 0} L} + B_0^l \ee^{-\I k_{{\rm II} 0} L} \right) \\
 + k_{{\rm II}  \pm 1} \left( B_{\pm 1}^r \ee^{\I k_{{\rm II} \pm 1} L} - B_0^l \ee^{-\I k_{{\rm II} \pm 1} L} \right),
\end{multline}
where we have already incorporated the zeroth order terms \eqref{eq:App_zero1}-\eqref{eq:App_zero4}. 

Solving for the transmission amplitudes in the side channels $C_{\pm 1}^r$ we are able to establish a distinct scaling behaviour. For when the scalar potential becomes opaque $\sqrt{\mu V_0}L \gg 1$, the exponential, suppressive behaviour dominates. Hence, the amplitude decreases over the barrier length
\begin{align}
 & \lvert C_{+1}^r \rvert^2 \approx \frac{8 \mu \chi_0^2 E}{V_0} (V_0 - E - \omega) \ee^{-2  k_{{\rm II} +1} L },\\
 & \lvert C_{-1}^r \rvert^2 \approx \frac{8 \mu \chi_0^2 E}{V_0} (V_0 - E) \ee^{-2  k_{{\rm II} 0} L}.
 \label{Eq:pert_C}
\end{align}

On the other hand, for easily transmissible barriers $\sqrt{\mu V_0}L \ll 1$ we find, under the constraint $E > \omega$, for the transmission amplitudes 
\begin{align}
 \lvert C_{\pm 1}^r \rvert^2 \approx \left( \frac{\mu V_0 L \chi_0 }{2} \right)^2 \left( \frac{ \sqrt{2\mu E}}{\sqrt{2 \mu (E \pm \omega)}} - 1 \right)^2. 
\end{align}
The transmitted probability currents 
\begin{align}
j^{\pm 1}_\mathrm{trans} = \left(1 \pm \frac{\omega}{E} \right) \lvert C_{\pm 1}^r \rvert^2,
\end{align}
then satisfy the relation
\begin{align}
\frac{j_\mathrm{trans}^{+1}}{j_\mathrm{trans}^{-1}} = \frac{ \sqrt{E-\omega} }{ \sqrt{E+\omega} }
\frac{ \left( \sqrt{E+\omega} - \sqrt{E} \right)^2}{ \left( \sqrt{E-\omega} - \sqrt{E} \right)^2 } < 1.
\end{align}
Hence, for a small rectangular barrier, the transmitted current through the first upper side channel is smaller than the first lower side channel.

Furthermore, both channels have at least one maximum as a function of barrier length $L$. The position of these maxima can be estimated as 
\begin{align}\label{max}
 L^{\pm 1}_\mathrm{max} \sim \frac{1}{ \sqrt{ 2\mu (V_0 - E \mp \omega)}}. 
\end{align}
As we see the position of this maximum is different for lower and upper channels, respectively.

\subsection{Opaque Barrier Approximation} 
\label{Sec:Rec3}

We start with a general ansatz for the transmitted wave function
\begin{equation}
 \psi (x,t) = \int {\rm d}E \, \psi_{\rm trans}(E) \ee^{-\I E t + \I \sqrt{2 \mu E} \big( x - \chi(t) \big) }, 
\end{equation}
with quiver motion $\chi(t)$. For an incoming test energy $E_{\rm in}$ we find within the opaque barrier approximation ($V_0 \gg \omega$, $V_0 \gg E_{\rm in}$, $ L \gg \chi_0$) for the Fourier coefficients 
\begin{multline}
 \psi_{\rm trans} (E) \approx 4 \sqrt{ \frac{E_{\rm in}}{V_0}} \ee^{-\sqrt{2 \mu V_0} L + _{\rm in} \mathfrak{T} } 
 \ee^{-\I \sqrt{2 \mu E} L} \\
 \times \int \frac{{\rm dt}}{2\pi} 
 \ee^{ \I (E - E_{\rm in}) t} 
 \ee^{ \sqrt{2 \mu V_0} \big( \chi(t+i\mathfrak{T}) - \chi(t) \big) },
\end{multline}
where in this specific case the Landauer-Büttiker time \cite{BL82} takes on the form
\begin{align}
 \mathfrak{T} = L \sqrt{ \frac{\mu}{2V_0} }. 
\end{align}
Due to the fact that the last term in the equations is again time-periodic we can apply a Fourier decomposition 
\begin{align}
 \ee^{ \sqrt{2 \mu V_0} \big( \chi(t + \I \mathfrak{T} ) - \chi(t) \big) }
 = \sum_m \phi_m \ee^{-\I m \omega t}, 
\end{align}
with weights
\begin{align}
 \phi_m & = \frac{\omega}{2\pi} \int_0^{\frac{2\pi} {\omega}} {\rm d}t \ee^{\I m \omega t} \ee^{ \sqrt{2 \mu V_0} \big( \chi( t + \I \mathfrak{T} ) - \chi(t) \big) } \notag \\
 & = \left( \frac{ \ee^{\omega \mathfrak{T} } - 1}{\ee^{-\omega \mathfrak{T} } - 1} \right)^ { \frac{m}{2} } \notag \\
 & \times I_m \left( \lvert \chi_0 \rvert \sqrt{ 2 \mu V_0 \left( \ee^{\omega \mathfrak{T}} - 1 \right) \left( \ee^{-\omega \mathfrak{T}} - 1 \right) }
 \right),
 \label{Eq:phi_m}
\end{align}
where $I_m$ are Bessel functions of the first kind. On this basis we can perform the time integration and write
\begin{equation}
 \psi_{\rm trans} (E) \approx \psi_{\rm trans}^0 \sum_m \delta(E - E_{\rm in} - m \omega) \phi_m,
 \label{Eq:psi_m}
\end{equation}
with the bare wavefunction
\begin{align}
 \psi_\mathrm{trans}^0= 4 \sqrt{ \frac{ E_{\rm in} }{V_0}} 
 \ee^{- \sqrt{ 2 \mu V_0} L + E_{\rm in} \mathfrak{T} } \ee^{- \I \sqrt{2 \mu E} L }.
\end{align}
Consequently, we find for the transmitted probability currents 
\begin{align}
j_\mathrm{trans}^m = \frac{k_m}{k_0} \lvert \psi_{\rm trans} \, \phi_m \rvert^2,
\end{align}
with $k_m = \sqrt{2 \mu \left( E_{\rm in} + m\omega \right) }$.

At this point a comparison with the perturbative approach, Sec. \ref{Sec:Rec2}, is in order. The Bessel function in Eq. \eqref{Eq:phi_m} evaluated for small arguments yields
\begin{multline}
 \phi_m \approx 
 \frac{ \big( \chi_0 \sqrt{2 \mu V_0} \big)^{\lvert m \rvert}}{2^{\lvert m \rvert } \, \lvert m \rvert !} \\
 \times \left( \ee^{ \omega \mathfrak{T}} - 1 \right)^{ \frac{m}{2} + \frac{\lvert m \rvert}{2}}
 \left( \ee^{-\omega \mathfrak{T}} - 1 \right)^{-\frac{m}{2} + \frac{\lvert m \rvert}{2}}.
\end{multline}
In the limit of large $L$ and $m=1$ (first upper sideband) we obtain $\phi_m \sim \ee^{m \omega \mathfrak{T}}$. The amplitude squared of Eq. \eqref{Eq:psi_m} already taking into account the delta-function thus scales as $\propto \ee^{-2L \left( \sqrt{2 \mu V_0} - (E + \omega) \sqrt{\mu/V_0} \right)}$. Correspondingly, for large $L$ and $m=-1$ (first lower sideband) we find a scaling of $\propto \ee^{-2L \left( \sqrt{2 \mu V_0} - E \sqrt{\mu/V_0} \right)}$. These results agree very well with the perturbatively calculated probabilities in Eq. \eqref{Eq:pert_C} provided that the momentum vector $k$ is expanded in a Taylor series.

\section{Truncated 1/r potential}
\label{Sec:AppTrunc}

%Additionally, we display the analytical solutions for tunneling through a truncated Coulomb potential in terms of Whittaker functions.
As we are also discussing nuclear fusion, that is essentially tunneling transmission through a Coulomb barrier, it is educational to present the analytical solution for static tunneling. Equations for field-assisted quantum tunneling can, in principal, also be derived in the same way.

Again, the starting point for the derivation is the Schr\"odinger equation which for a general Coulomb potential takes on the form
\begin{equation}
 \left( -\frac{1}{2 \mu} \frac{{\rm d}^2}{{\rm d}x^2} + \frac{\alpha}{x} \right) \psi(x) = E \psi(x),
\end{equation}
where the fine-structure constant $\alpha=(q^2 q_1 q_2)/(4 \pi \varepsilon_0)$ controls the strength of the potential. In the context of nuclear fusion this control parameter would take on the value for the fine structure constant $\alpha$ times a charge factor. The solutions for tunneling and reflection amplitudes for a particle of energy $E$ in an $1/r$-potential is given in terms of Whittaker functions ${\cal W}$. Specifically, we have for the wave function at the inner turning point $r_0$, that is where we truncate the $1/r$-potential to keep the potential finite, 
\begin{equation}
 \psi(x) = c_E^\ast v_E^\ast(x) + r_E c_E v_E(x),    
\end{equation}
with functions 
\begin{align}
 &v_E(x) &&=&& {\cal W} \left(-\I V_0 \sqrt{\frac{\mu}{2E}}, 1/2, -2\I \sqrt{2 \mu E} x \right),\\
 &c_E &&=&& \left(-2 \I \sqrt{2 \mu E} \right)^{ \I V_0 \sqrt{\mu /2E} },
\end{align}
and reflection coefficient
\begin{equation}
 r_E = \left. -\frac{c_E^\ast}{c_E} \frac{ \dfrac{{\rm d} v_E^\ast(x)}{{\rm d}x} +\I \sqrt{2 \mu E} v_E^\ast(x) }{ \dfrac{{\rm d} v_E(x)}{{\rm d}x} +\I \sqrt{2 \mu E} v_E(x) } \right\vert_{x=r_0}.
\end{equation}
The transmission coefficient $t_E$ is accordingly given by
\begin{equation}
 t_E = \Big. c_E^\ast v_E^\ast(x) + r_E(x) c_E v_E(x)  \Big\vert_{x=r_0}.
\end{equation}

%%%%%%%%%%%%%%%%%%%%%%%%%%%%%%%%%%%%%%%%%%%%%%%%%%%%%%%%%%%%%%%%%%%%%%%%%%%%%%%%
%%%%%%%%%%%%%%%%%%%%%%%%%%%%%%%%%%%%%%%%%%%%%%%%%%%%%%%%%%%%%%%%%%%%%%%%%%%%%%%%
\section{Supplemental Figures} 
\label{app:D}
%%%%%%%%%%%%%%%%%%%%%%%%%%%%%%%%%%%%%%%%%%%%%%%%%%%%%%%%%%%%%%%%%%%%%%%%%%%%%%%%
%%%%%%%%%%%%%%%%%%%%%%%%%%%%%%%%%%%%%%%%%%%%%%%%%%%%%%%%%%%%%%%%%%%%%%%%%%%%%%%%

Finally, we would like to present additional sets of data in the form of auxiliary figures. These are used to provide additional context to our discussions, as they allow us to highlight certain details that are less suited for a general audience.

\subsection{Rectangular barrier} 

%%%%%%%%%%%%%%%%%%%%%%%%%%%%%%%%%%%%%%%%%%%%%%%%%%%%%%%%%%%%%%%%%%%%%%%%%%%%%%%%
\begin{figure}[t]
 \includegraphics[trim={0cm 0.0cm 0cm 0.cm},clip,width = 0.49\textwidth]{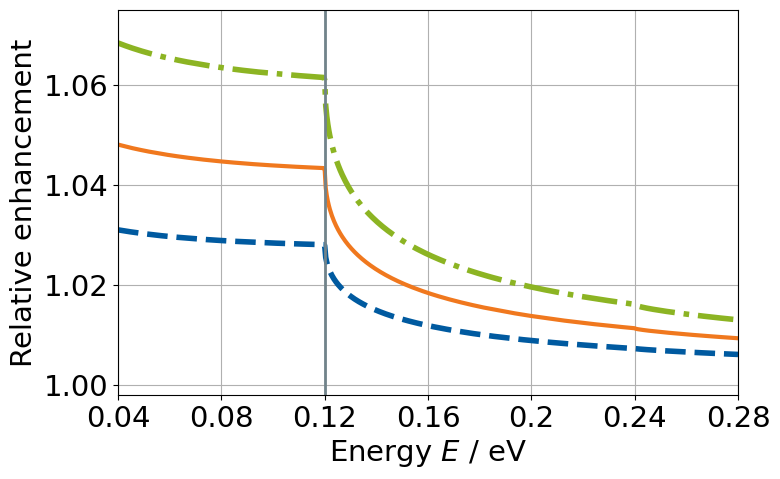}
 \caption{
 Relative enhancement in the electron transmission probability through a rectangular barrier as a function of initial energy $E$ for different external field strengths $\mathfrak{E}_0$ ($1.6 \times 10^{8}$ V/m for the blue dashed line, $2.0 \times 10^{8}$ V/m for the orange solid line and $2.4 \times 10^{8}$ V/m for the green dot-dashed line). The quiver frequency is fixed to $\omega = 0.12$ eV, the barrier height is $V_0=6$ eV, the barrier width $L=0.2$ nm and the asymptotic potential behind the barrier is $V_1 = 352 $ keV. 
 }
 \label{fig:B6}
\end{figure}
%%%%%%%%%%%%%%%%%%%%%%%%%%%%%%%%%%%%%%%%%%%%%%%%%%%%%%%%%%%%%%%%%%%%%%%%%%%%%%%%

In Fig.~\ref{fig:B6} the relative enhancement in the total tunneling probability is shown for strongly attractive asymptotic potential. If the initial energy $E$ is higher than the quiver frequency $\omega$ the relative enhancement diminishes. 

\subsection{Coulomb barrier} 

\begin{figure}[t]
 \includegraphics[trim={0cm 0.0cm 0cm 0.cm},clip,width = 0.48\textwidth]{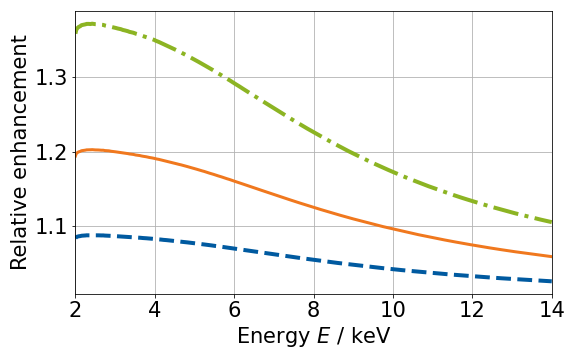}
 \includegraphics[trim={0cm 0.0cm 0cm 0.cm},clip,width = 0.48\textwidth]{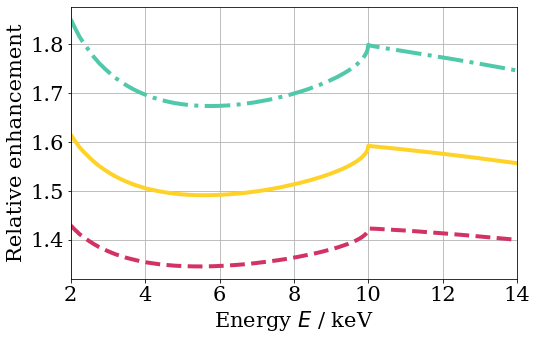} 
 \caption{
 Relative enhancement in particle transmission probability in a truncated Coulomb potential with $V_1=0$ as a function of initial energy $E$ for frequency $\omega = 2$ keV (top) as well as $\omega=10$ keV (bottom). Three different field strengths have been used, respectively. For the plot at the top we have ${\mathfrak E}_0 = 1.0 \times 10^{16}$ V/m for the blue dashed line, ${\mathfrak E}_0=1.5 \times 10^{16}$ V/m for the orange solid line and $2.0 \times 10^{16}$ V/m for the green dot-dashed line. For the plot at the bottom we have ${\mathfrak E}_0 = 3.0 \times 10^{17}$ V/m for the magenta dashed line, ${\mathfrak E}_0=3.5 \times 10^{17}$ V/m for the yellow solid line and $4.0 \times 10^{17}$ V/m for the cyan dot-dashed line.
 An overall increase in the transmission probability the smaller the initial energy is chosen is accompanied by a resonant amplification when $E - \omega \approx 0$. In all plots the asymptotic potential levels have been fixed to zero.
 }
 \label{fig:AppAC2}
\end{figure}

In Fig.~\ref{fig:AppAC2} a Coulomb potential with equal asymptotic levels moving laterally gives rise to resonant behaviour in the relative tunneling probability. These plots accompany the figure in the main text, cf. Fig.~\ref{fig:AC2}, showing that the peak position indeed varies with the quiver frequency $\omega$.

\begin{figure}[t]
 \includegraphics[width=0.49\textwidth]{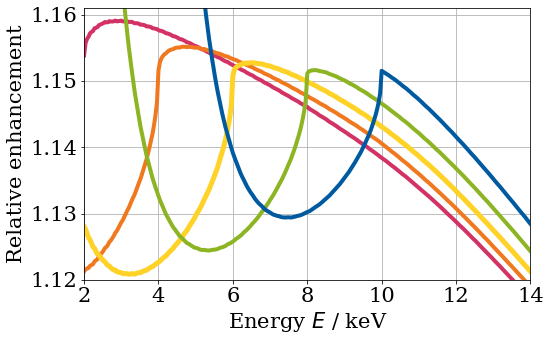}
 \caption{
 Relative enhancement in the tunneling probability as a function of the initial energy $E$ and for various potential depths $V_1$ in a truncated Coulomb potential. Peaks in the curves coincide with $E + n\omega = -V_1$ for $n < 0$. The quiver frequency is $\omega=6$ keV. The field strength was fixed at ${\mathfrak E}_0 \approx 8 \times 10^{16}$ V/m. Potential depths are color-coded as $-V_1=4$ keV (blue), $-V_1=2$ keV (green), $-V_1=0$ keV (yellow), $-V_1=4$ keV (orange) and $-V_1=-4$ keV (magenta).
 }
 \label{fig:AppAC5}
\end{figure}

In Fig.~\ref{fig:AppAC5} the relative enhancement in tunneling transmission is displayed as a function of the initial energy but for varying asymptotic potential levels. The resonance structure appears if $E = n\omega + V_1$, with particle energy $E$, potential height $V_1$ and frequency $\omega$. If $V_1$ is smaller than the potential at the side of the incoming particle the resonances are shifted to the left to smaller energies. If, on the other hand, the potential features a level $V_1 > 0$ the curve shifts to the right. Note that there is another resonance when the initial energy coincides with $V_1$. But here the relative enhancement is too strong to plot so we decided to only show the peaks at $n=1$.

\begin{figure}[h]
 \includegraphics[width = 0.49\textwidth]{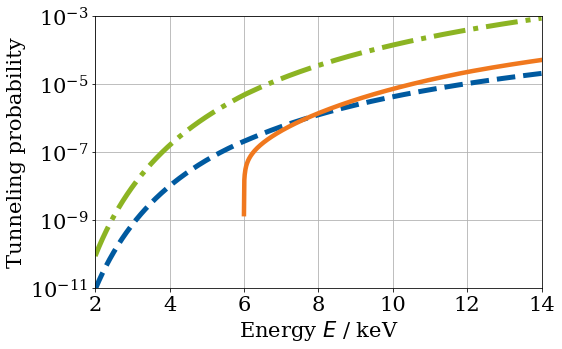}
 \caption{
 Log-linear plot of the tunneling probability in a truncated Coulomb potential as a function of the initial energy $E$ displayed in terms of main channel (green, dot-dashed) as well as first upper (blue) and first lower (orange) side channel. 
 The field frequency is $\omega=6$ keV and the field strength ${\mathfrak E}_0 = 10^{17}$ V/m. Lower energy sidebands can only contribute if their final energy is above the potential's asymptotic level, here $V_1 = 0$.
 }
 \label{fig:AppC1}
\end{figure}

In Fig.~\ref{fig:AppC1} we display the total transmission probability in a Coulomb potential as a function of a particle's initial energy for various field strengths. Overall, having an electric field boosts the transmission probability which, in this specific case, grows monotonically for all three energy channels. Nevertheless, there is an obvious difference in scaling behaviour especially for the two sidebands $E-\omega$ and $E+\omega$, as a particle's final net energy has to be strictly positive.

%%%%%%%%%%%%%%%%%%%%%%%%%%%%%%%%%%%%%%%%%%%%%%%%%%%%%%%%%%%%%%%%%%%%%%%%%%%%%%%%
%%%%%%%%%%%%%%%%%%%%%%%%%%%%%%%%%%%%%%%%%%%%%%%%%%%%%%%%%%%%%%%%%%%%%%%%%%%%%%%%

%\textcolor{white}{x} %Makes things work!
%\vspace{2cm}
%\clearpage

%%%%%%%%%%%%%%%%%%%%%%%%%%%%%%%%%%%%%%%%%%%%%%%%%%%%%%%%%%%%%%%%%%%%%%%%%%%%%%%%
%%%%%%%%%%%%%%%%%%%%%%%%%%%%%%%%%%%%%%%%%%%%%%%%%%%%%%%%%%%%%%%%%%%%%%%%%%%%%%%%

%\newpage

%%%%%%%%%%%%%%%%%%%%%%%%%%%%%%%%%%%%%%%%%%%%%%%%%%%%%%%%%%%%%%%%%%%%%%%%%%%%%%%%
%%%%%%%%%%%%%%%%%%%%%%%%%%%%%%%%%%%%%%%%%%%%%%%%%%%%%%%%%%%%%%%%%%%%%%%%%%%%%%%%

%\vspace{-1cm}
%\textcolor{white}{x} %Makes things work!
%\newpage

\clearpage

%%%%%%%%%%%%%%%%%%%%%%%%%%%%%%%%%%%%%%%%%%%%%%%%%%%%%%%%%%%%%%%%%%%%%%%%%%%%%%%%
%%%%%%%%%%%%%%%%%%%%%%%%%%%%%%%%%%%%%%%%%%%%%%%%%%%%%%%%%%%%%%%%%%%%%%%%%%%%%%%%
%\paragraph{Bibliography.--} 
%%%%%%%%%%%%%%%%%%%%%%%%%%%%%%%%%%%%%%%%%%%%%%%%%%%%%%%%%%%%%%%%%%%%%%%%%%%%%%%%
%%%%%%%%%%%%%%%%%%%%%%%%%%%%%%%%%%%%%%%%%%%%%%%%%%%%%%%%%%%%%%%%%%%%%%%%%%%%%%%%
%\clearpage
\interlinepenalty=10000

\bibliographystyle{unsrt}

\end{document}